\documentclass[a4paper,11pt]{article}
\usepackage{jcappub}
\usepackage[utf8]{inputenc}
\usepackage{amsmath,mathtools}

\usepackage{tensor}
\usepackage{amssymb}
\usepackage{graphicx}
\usepackage{tikz}
\usepackage{physics}
\usepackage{lipsum}  
\usepackage{bigints}
\usepackage{verbatim}
\usepackage{bbold}
\usepackage[normalem]{ulem}

\newcommand{\leri}[1]{\left(#1\right)}
\newcommand{\lerisq}[1]{\left[#1\right]}
\newcommand{\jor}{\frac{1}{2\kappa^2}\int d^4x\,\sqrt{-g}\;}
\newcommand{\g}{\sqrt{-g}}
\newcommand{\mat}{\mathcal{L}_m}

\title{Gravitational waves in Palatini gravity for a non-minimal geometry-matter coupling}

\author[a,1]{F. Bombacigno,\note{Corresponding author.}}
\author[b]{F. Moretti,}
\author[a,c]{Gonzalo J. Olmo,}

\affiliation[a]{Instituto de Física Corpuscular (IFIC), CSIC‐Universitat de València, Spain}
\affiliation[b]{Fusion and Nuclear Safety Department, ENEA, C. R. Frascati, Via E. Fermi 45, 00044 Frascati, Italy}
\affiliation[c]{Universidade Federal do Cear\'a (UFC), Departamento de F\'isica, Campus do Pici, Fortaleza - CE, C.P. 6030, 60455-760 - Brazil.}

\emailAdd{flavio2.bombacigno@uv.es}
\emailAdd{fabio.moretti.1@enea.it}
\emailAdd{gonzalo.olmo@uv.es}

\abstract{We discuss the propagation of gravitational waves over a non-Riemannian spacetime, when a non-minimal coupling between the geometry and matter is considered in the form of contractions of the energy momentum tensor with the Ricci and co-Ricci curvature tensors. We focus our analysis on perturbations on a Minkowski background, elucidating how derivatives of the energy momentum tensor can sustain non-trivial torsion and non-metricity excitations, eventually resulting in additional source terms for the metric field. These can be reorganized in the form of D'Alembert operator acting on the energy momentum tensor and the equivalence principle can be reinforced at the linear level by a suitable choice of the parameters of the model. We show how tensor polarizations can exhibit a subluminal phase velocity in matter, evading the constraints found in General Relativity, and how this allows for the kinematic damping in specific configurations of the medium and of the geometry-matter coupling. These in turn define regions in the wavenumber space where propagation is forbidden, leading to the appearance of typical cut-off scale in the frequency spectrum.}
\makeatletter
\gdef\@fpheader{}
\makeatother
\begin{document}

\maketitle
\flushbottom
\section{Introduction}\label{sec: 1}
\noindent Modified theories of gravity have received a constant growing attention over the past decades, at the urging of astrophysical and cosmological phenomena which are not likely to be explained within the theoretical mark of General Relativity (GR). We refer in particular to the problem of dark matter \cite{Feng:2010gw,Bertone:2016nfn,Cirelli:2024ssz} and late-time accelerated expansion of the Universe \cite{SupernovaCosmologyProject:1998vns,SupernovaSearchTeam:1998fmf}, which have been recently complemented by the so-called $H_0$ tension \cite{DiValentino:2021izs,Dainotti:2021pqg,CosmoVerse:2025txj} and the dynamical dark energy issue \cite{DESI:2024mwx,DESI:2025zgx}. From this perspective, it is therefore of primary interest to establish phenomenological signatures that could facilitate the comparison between observations and predictions. 
\\ At the present day, one of the most valuable strategy to test alternative theories relies on gravitational wave astronomy. It is indeed expected that third generation detectors (Cosmic Explorer and Einstein Telescope) will be sensitive to mergers of black holes up to a redshift $z\simeq30$ \cite{Ng:2020qpk}, also increasing our resolution power for different polarizations \cite{Takeda:2019gwk,Isi:2022mbx}. As a matter of fact, tensor, vector, and scalar modes are affected in different ways by the material media they traverse, and an attenuation of the gravitational signal can arise in the presence of a variety of scenarios, e.g. dissipative fluids \cite{Hawking:1966qi,Madore,Madore:1972ww,Prasanna:1999pn,Anile}, an expanding Universe \cite{1978SvA....22..528Z,Weinberg:2003ur,Flauger:2017ged} and a cosmological neutrino background \cite{Lattanzi:2005xb,Lattanzi:2010gn,Benini:2010zz}. In this work, we are mainly interested into the Landau damping, comprising the damping of the wave perturbation as a result of its interaction with a non-collisional medium. In the context of kinetic theory, non-collisional refers to the fact that in the Boltzmann equation, describing the evolution of the distribution function for the medium, the collision integral can be neglected, so that we are left with an Einstein-Vlasov system once the equation for the metric perturbation is included. The feasibility of this phenomenon in GR has been investigated for a flat Minkowski or an expanding FLRW background in \cite{Chesters:1973wan,PhysRevD.13.2724,Gayer:1979ff,Weinberg:2003ur,Lattanzi:2005xb,Benini:2010zz,Flauger:2017ged,Baym:2017xvh}. These studies have elucidated that tensor polarizations in GR can suffer Landau damping only when anisotropies are included in the background medium, or the coupling between wave perturbations and curvature is considered. Landau damping is indeed strictly forbidden for isotropic configurations and static flat spacetimes, where  gravitational modes are characterized by a superluminal phase velocity in the medium, which represents the sufficient condition for the non-existence of a similar phenomenon. These results suggest that in order the Landau damping to actually occur on a Minkowski spacetime, a non-GR equation of motion must onset at linear level, possibly exhibiting either non standard differential operators acting on the wave polarizations or a set of modified source terms. These non-Einsteinian contributions are expected to give rise in matter to deformed dispersion relations, where the dependence of the frequency on the wave number and the theory parameters is enclosed.
\\ In a series of previous works we discussed the impact of modified wave operators on Landau damping viability, focusing on the propagation of gravitational waves in Horndeski \cite{Moretti:2020kpp,Moretti:2022xem} and parity violation \cite{Bombacigno:2022naf} theories of gravity. We showed that the massive scalar mode contained in Horndeski models can experience Landau damping, when a characteristic relation between the mass of the scalar perturbation and the thermodynamic properties of the medium is satisfied. We then extended these results to parity violation scenarios, and we proved that in the presence of gravitational birefringence, tensor polarizations can also suffer Landau damping, provided specific regions of the wavenumber space are considered, ultimately resulting in a typical frequency band structure for the transmitted gravitational signal. 
\\In this work, we investigate the second possibility, and we examine whether Landau damping can take place or not when the canonical d'Alembert operator is still acting on tensor polarizations, but in the presence of non-standard source terms. Non-standard here denotes contributions that are known to arise in the presence of a non-minimal coupling between matter and geometry, as it occurs when the gravitational action is enlarged to include also a dependence on matter derived quantities. We refer in particular to a broad class of theories whose Lagrangian density is of the form $f(R,T)$ or $f(R,\mathcal{L}_m)$ \cite{Harko:2010mv,Harko:2011kv,Harko:2012hm}, with $T$ and $\mathcal{L}_m$ standing for the trace of the energy momentum tensor and the matter Lagrangian density, respectively. In the last years these models have received great attention, for their cosmological and astrophysical implications \cite{Harko:2014gwa,Harko:2014aja,Wu:2018idg,Jaybhaye:2022gxq,BarrosoVarela:2024htf}, being possible to reconcile the geometry-matter coupling in terms of thermodynamical open systems, featured by matter creation during irreversible processes \cite{Harko:2014pqa,Harko:2015pma}. Such an interpretation is possible since in these theories the energy momentum tensor is not covariantly conserved, leading to non-geodesic motion for test particles, where an extra force orthogonal to the 4-velocity usually appears \cite{Bertolami:2007gv}. These predictions are consistent with an explicit violation of the equivalence principle, and theories can be potentially constrained by solar system experimental observations. Further generalizations have been also formulated, involving the contraction of the energy momentum tensor with itself, like in the so-called energy momentum squared gravity $f(R,T_{\mu\nu}T^{\mu\nu})$ \cite{Board:2017ign,Cipriano:2023yhv,Rosa:2023guo}, or with the Ricci curvature tensor, leading to the $f(R,R_{\mu\nu}T^{\mu\nu})$ models \cite{Haghani:2013oma,Odintsov:2013iba,Sharif:2013kga,Feng:2025zau}. In this respect, the coupling $R_{\mu\nu}T^{\mu\nu}$ is of particular interest for its dynamical implications, being possible for Dolgov-Kawasaki instabilities to show up in cosmological settings \cite{Haghani:2013oma}, or ghostlike modes to be excited \cite{Ayuso:2014jda}. As outlined in \cite{Afonso:2017bxr,Barrientos:2018cnx,Fox:2018gop}, these limitations can be potentially overcome by embracing a different perspective about the geometric structure of the theory, allowing for the affine connection to be an independent dynamical entity with respect to the metric field. More than on a pure metric-affine formalism, these works rely on a Palatini approach, where connection only enters the definition of curvature-derived quantities and does not directly couples to matter fields. In addition, they also assume a symmetric connection, setting a priori torsion to zero. Under these hypotheses, a solution for the connection can be still obtained, resulting in the Levi-Civita connection of an auxiliary, energy momentum–dependent metric, related to the original metric by a (matrix) disformal transformation. Besides the discussion in \cite{Barrientos:2018cnx,Afonso:2017bxr,Fox:2018gop}, the coupling of geometry with matter in a metric-affine (Palatini) framework has been rarely addressed, and a comprehensive analysis about the role of the connection in the presence of both torsion and nonmetricity is still missing. In this work we move towards such a direction, and we present the first (approximated) solution known in the literature for a generic connection with no symmetry restrictions. Moreover, being the final goal of our investigation a detailed analysis of the propagation of gravitational waves in the weak-field limit, we choose to restrict our attention to geometry-matter couplings which are linear both in the energy momentum tensor and in the curvature. They are indeed the only contributions to survive when linearization on a Minkowski background is performed. We note that in a general metric-affine setting the coupling $R_{\mu\nu}T^{\mu\nu}$ is not the only term complying with these requirements, since other contractions built out of different Riemann tensor components are actually feasible. That leads us to consider also the contribution $\tilde{R}_{\mu\nu}T^{\mu\nu}$, with $\tilde{R}_{\mu\nu}$ denoting the so-called co-Ricci tensor (see Sec.~\ref{sec: 2} for technical details). Interestingly, these terms are also the only combinations compatible with projective invariance, the symmetry associated with the invariance under vector translation of the affine connection, whose violation is often related to dynamical instabilities in metric-affine scenarios \cite{BeltranJimenez:2019acz,BeltranJimenez:2020sqf,Iosifidis:2018jwu,Iosifidis:2019fsh,Sauro:2022hoh}. Furthermore, similar terms guarantee that the equation for the connection is still both algebraic and linear in the torsion and non metricity, with differential operators appearing only on the energy momentum tensor components. In particular, we chose to study our model in a multi-scalar-tensor form, in analogy with the standard treatment of Palatini $f(R)$ gravity \cite{Olmo:2005hc,Olmo:2011uz}. In order to avoid the ambiguities due to the possible self-recursive definition of the energy momentum tensor, when the gravitational sector is accompanied with the matter action, we adopt the perspective of \cite{Ayuso:2014jda}. The energy momentum tensor is here defined solely from the action for the matter fields, and only then entering the gravitational Lagrangian. Then, even if such an operational energy momentum tensor is not covariantly conserved as in the metric case, we show that at least at the linear level the parameters of the theory can be still set in order to partially recover conservation. In particular, that can be achieved by simply neglecting in the linearized theory the contribution from the Ricci tensor, so that the non-minimal coupling of matter with geometry is now completely sustained by the co-Ricci tensor. Under these assumptions and after some quite technical and involved manipulations, the linearized equation for the connection can be then solved exactly. It is therefore possible to obtain the effective equation for the metric perturbation, where the final effect of the non-minimal coupling between geometry and matter just results in an additional d'Alembert operator acting on the energy momentum tensor. This raises the issue, as in the metric case, of potential dynamical instabilities triggered by higher order equations of motion, due to the presence in $T_{\mu\nu}$ of first derivatives of the matter fields. In this work we do not analyse in detail such an important aspect and we remind the reader to \cite{Ayuso:2014jda} for an exhaustive and critical discussion on the topic. We just observe that such a situation is quite similar to what occurs in ordinary Palatini $f(R)$ gravity, where covariant derivatives acting on the scalaron of the Jordan frame, depending on the trace of the energy momentum tensor, appear in the equation for the metric field without resulting in dynamical instabilities.
\\Here, we focus on the possible phenomenological implications on the propagation of gravitational waves\footnote{For analysis concerning the propagation of gravitational waves in Palatini and geometry-matter coupling frameworks see for example \cite{BeltranJimenez:2015zhq,Bertolami:2017svl,Varela:2024egg,Lopez:2025gfu}.}, when the mutual interplay between matter and tensor modes is examined from the perspective of the kinetic theory. Building on our previous results in \cite{Moretti:2020kpp,Moretti:2022xem,Bombacigno:2022naf}, we then investigate the feasibility and the properties of the Landau interaction for tensor polarizations. Our main finding is the existence of a set of modified dispersion relations for the frequency, exhibiting characteristic cut-off scales in the Fourier momentum space. These wave-numbers are inherently related to the coefficient of the coupling between the co-Ricci tensor and the matter, as well as to the thermodynamic properties of the medium the gravitational signal is traveling through. In particular, these scales define where the propagation is forbidden or allowed, and should be the latter the case, where the Landau damping can take place. Then, as outlined in \cite{Avelino:2012qe,BeltranJimenez:2017doy}, by demanding the resulting modified Poisson equation be within the observational constraint, we are able to determine a lower bound for the momentum cut-off. Eventually, this guides us in discussing the possible phenomenological implications on current and future interferometer observations.
\\The paper is organized as follows. In Sec.~\ref{sec: 2} we present the generic properties of the model, elucidating the role of the different possible couplings for a metric-affine geometry. In Sec.~\ref{sec: 3} we derive the complete set of equations for the connection, the metric and the scalar fields. In Sec.~\ref{sec: 4} we provide a detailed discussion of the linearized theory, by offering a step by step derivation of the solution for torsion and non-metricity, as well as of the effective equation for the metric perturbation. In Sec.~\ref{sec: 5} we reformulate the problem of the propagation of gravitational waves in matter within the context of kinetic theory, and we establish the different region in the wave-number space where the propagation is allowed or forbidden, and Landau damping can appear or not. In Sec.~\ref{sec: 6} we determine the phenomenological constraints for realistic scenarios where the assumption of the model are satisfied. In Sec.~\ref{sec: 7} we draw our conclusions, and in  App.~\ref{app: a} we give an overview of the basic concepts and geometric quantities we use throughout the paper for the metric-affine framework.
\\Gravitational coupling is set as $\chi=8\pi$, using geometrized units $G=c=1$. Boltzmann constant is set to unity, i.e. $k_B=1$.

\section{The model}\label{sec: 2}
\noindent In this work we consider a Palatini formulation of the model introduced in \cite{Ayuso:2014jda}, that here we extend with the aim of including the additional couplings between geometry and matter which are compatible with the symmetries of the non-Riemannian spacetime. This amounts to discuss the action
\begin{equation}
    S=\jor \leri{f(R,X,Y)+2\kappa^2\mat(g;\psi_i)},
    \label{eq: action}
\end{equation}
where the scalar quantities $X,Y$ read respectively as:
\begin{equation}
X\equiv R_{\mu\nu}T^{\mu\nu},\quad Y\equiv g^{\mu\nu}R\indices{^\rho_{\mu\sigma\nu}}T\indices{_\rho^\sigma}.
    \label{eq: def X Y}
\end{equation}
Here the Riemann tensor is defined in terms of the independent connection as
\begin{equation}
    R\indices{^\rho_{\mu\sigma\nu}}=\partial_\sigma\Gamma\indices{^\rho_{\mu\nu}}-\partial_\nu\Gamma\indices{^\rho_{\mu\sigma}}+\Gamma\indices{^\rho_{\tau\sigma}}\Gamma\indices{^\tau_{\mu\nu}}-\Gamma\indices{^\rho_{\tau\nu}}\Gamma\indices{^\tau_{\mu\sigma}},
\end{equation}
and the energy momentum tensor is given by
\begin{equation}
    T_{\mu\nu}\equiv -\frac{2}{\sqrt{-g}}\frac{\delta (\sqrt{-g}\mat)}{\delta g^{\mu\nu}}=g_{\mu\nu}\mat-2\frac{\delta\mat}{\delta g^{\mu\nu}},
    \label{eq: def Tmn}
\end{equation}
with $\mat$ the Lagrangian of the matter fields denoted collectively by $\psi_i$. 
We remark that in the standard metric formulation the only nontrivial contraction of the energy momentum tensor with the Riemann tensor is determined by the scalar $X$, while in a metric-affine setting there are more options. In particular, one can consider the homothetic curvature and the co-Ricci tensor, displayed respectively by $\hat{R}_{\mu\nu}=R\indices{^\rho_{\rho\mu\nu}}$ and $\tilde{R}_{\mu\nu}=g^{\rho\sigma}R_{\mu\rho\nu\sigma}$. The homothetic curvature is the other possible trace for the Riemann tensor when non-metricity is nonvanishing (see App.~\ref{app: a}), but since it is completely antisymmetric, it gives no contribution when contracted with the energy momentum tensor. This implies that the only nontrivial contraction one can consider beyond the Ricci tensor is given by the co-Ricci tensor, which is not a pure trace with respect to the former, but involves the metric in its construction. Interestingly, we note that scalars $X$ and $Y$ play a privileged role also when metric-affine symmetries, like projective invariance, are considered \cite{Sauro:2022hoh}. A projective transformation for the independent connection is, in general, given by the vector translation
\begin{equation}
\Gamma\indices{^\rho_{\mu\nu}}\rightarrow\tilde{\Gamma}\indices{^\rho_{\mu\nu}}=\Gamma\indices{^\rho_{\mu\nu}}+\delta\indices{^\rho_\mu}\xi_\nu,
    \label{eq: projective}
\end{equation}
which results in the Riemann tensor transformation rule
\begin{equation}
    \tilde{R}\indices{^\rho_{\mu\sigma\nu}}=R\indices{^\rho_{\mu\sigma\nu}}+\delta\indices{^\rho_\mu}\leri{\partial_\sigma\xi_\nu-\partial_\nu\xi_\sigma}.
\end{equation}
It follows that even if the Ricci and the co-Ricci tensor are not themselves invariant for a projective transformation, they are only affected in their antisymmetric part, so that when contracted with $T_{\mu\nu}$, the resulting scalars $X$ and $Y$ are left unchanged. This implies that neglecting higher order non-minimal couplings, involving a larger number of curvature and energy momentum tensors, the only combinations preserving projective invariance are the quantities $X$ and $Y$.
\\ Following the suggestion of \cite{Ayuso:2014jda}, it is convenient to rearrange \eqref{eq: action} in the equivalent scalar tensor form
\begin{equation}
    S=\jor \leri{\phi R+\chi_1 X+\chi_2 Y-W(\phi,\chi_1,\chi_2)+2\kappa^2\mathcal{L}_m(\psi_i,g)},
    \label{eq: action scalartensor}
\end{equation}
with $\phi=f_R,\;\chi_1=f_X,\;\chi_2=f_Y$ and the potential is defined as
\begin{equation}
    W=\phi R+\chi_1 X+\chi_2 Y-f(\phi,\xi,\chi_1,\chi_2),
\end{equation}
where the quantities $R,X,Y$ must be understood as evaluated in terms of $(\phi,\chi_1,\chi_2)$, with the inversion holding for a non-degenerate Hessian matrix $H_{ij}=\frac{\partial^2 f}{\partial a_i \partial a_j}$, with $a_i=R,X,Y$. 

The aim of this work is to elucidate how a non-trivial geometry-matter coupling can affect the propagation of GWs through the matter, when a Palatini formalism is considered. Under the hypothesis of a vanishing hypermomentum \cite{Iosifidis:2021nra,Iosifidis:2023but,Andrei:2024vvy}, corresponding to a matter Lagrangian not depending on the connection, we see that the variation of \eqref{eq: action scalartensor} with respect to $\Gamma\indices{^\rho_{\mu\nu}}$ can possibly result in energy momentum tensor contributions into the connection equation, potentially leading in the effective metric equation to derivative terms acting on the matter source (see Secs.~\ref{sec: 3}-\ref{sec: 4} below for the technical details). Based on the results of \cite{Bombacigno:2022naf}, we expect that in a kinetic theory description of the medium, this might induce a typical band structure in the GW spectrum, presumably modulated at linearized level by the background value of the fields $\chi_{1,2}$. 

\section{The equation for the connection, the metric, and the scalar fields}\label{sec: 3}
We start the discussion by performing the variation of \eqref{eq: action scalartensor} with respect to the connection, with the idea to relate the different components of $\Gamma\indices{^\rho_{\mu\nu}}$, i.e. torsion and non-metricity, to both the scalar fields and the energy momentum tensor. In doing so, we re-iteratively resort to the Palatini identity \eqref{eq: pal identity} for the variation of the Riemann tensor $\delta R\indices{^\rho_{\mu\sigma\nu}}$ and to the properties of the boundary terms for a metric-affine framework \eqref{eq: boundary term affine}, as they are introduced in App.~\ref{app: a}. To begin, we point out that the variation of the terms proportional to $\phi$ and $\chi_1$ can be collectively rewritten as
\begin{equation}
    \jor \leri{\phi g^{\mu\nu}+\chi_1 T^{\mu\nu}}\delta R_{\mu\nu},
\end{equation}
which by the application of \eqref{eq: pal identity} and \eqref{eq: boundary term affine} results in 
\begin{align}
    \tensor[^{(1)}]{\mathcal{E}}{_\lambda^{\mu\nu}}=&-\nabla_\lambda \leri{\g \leri{\phi g^{\mu\nu}+\chi_1 T^{\mu\nu}}}+\delta\indices{^\nu_\lambda}\nabla_\rho\leri{\g\leri{\phi g^{\mu\rho}+\chi_1 T^{\mu\rho}}}+\\
    &+\g\leri{\leri{\phi g^{\mu\nu}+\chi_1 T^{\mu\nu}}t_\lambda-\leri{\phi g^{\mu\rho}+\chi_1 T^{\mu\rho}}\leri{\delta\indices{^\nu_\lambda}t_\rho-t\indices{^\nu_{\rho\lambda}}}}.
    \label{eq: var conn 1}
\end{align}
We remind that here $\nabla$ stands for the covariant derivative built on the entire independent connection, which is not in general metric compatible. The tensor $t\indices{^\rho_{\mu\nu}}$ and $t_\mu=t\indices{^\rho_{\mu\rho}}$ denotes respectively the torsion and its trace. The second contribution is obtained from the variation of the term proportional to the scalar field $\chi_2$ and reads
\begin{equation}
    \jor \chi_2\; g^{\mu\nu}T\indices{_\rho^\sigma}\;\delta R\indices{^\rho_{\mu\sigma\nu}} \ ,
\end{equation}
leading to
\begin{align}
     \tensor[^{(2)}]{\mathcal{E}}{_\lambda^{\mu\nu}}=&-\nabla_\rho \leri{\g \chi_2 \,g^{\mu\nu}T\indices{_\lambda^\rho}}+\nabla_\rho\leri{\g\chi_2\, g^{\mu\rho}T\indices{_\lambda^\nu}}+\\
    &+\g\,\chi_2\leri{g^{\mu\nu}T\indices{_\lambda^\rho}t_\rho-T\indices{_\lambda^\nu}t^\mu+T\indices{_\lambda^\rho}t\indices{^{\nu\mu}_\rho}}.
    \label{eq: var conn 2}
\end{align}
It can be immediately checked that for $\mu=\lambda$ the identities $\tensor[^{(1)}]{\mathcal{E}}{_\mu^{\mu\nu}}=0=\tensor[^{(2)}]{\mathcal{E}}{_\mu^{\mu\nu}}$ hold, which display nothing but the invariance of the theory under projective transformations. Such a symmetry allows us to gauge away one vector component of the connection (see \cite{Bejarano:2019zco,Afonso:2017bxr,Iosifidis:2018zjj}), which for the sake of convenience we choose to be  the trace of the torsion tensor,  $t^\mu$ (see App.~\ref{app: a}). For this assumption, the equation for the connection simplifies as
\begin{align}
\tensor[^{(1)}]{\mathcal{E}}{_\lambda^{\mu\nu}}+\tensor[^{(2)}]{\mathcal{E}}{_\lambda^{\mu\nu}}=&-\nabla_\lambda \leri{\g \leri{\phi g^{\mu\nu}+\chi_1 T^{\mu\nu}}}+\\
&+\nabla_\rho\leri{\g\leri{\leri{\phi g^{\mu\rho}+\chi_1 T^{\mu\rho}}\delta\indices{^\nu_\lambda}-\chi_2 \leri{ \,g^{\mu\nu}T\indices{_\lambda^\rho}-g^{\mu\rho}T\indices{_\lambda^\nu}}}}+\nonumber\\
&+\g\leri{\leri{\phi g^{\mu\rho}+\chi_1 T^{\mu\rho}}t\indices{^\nu_{\rho\lambda}}+\chi_2\,T\indices{_\lambda^\rho}t\indices{^{\nu\mu}_\rho}}\nonumber,
\end{align}
which by using the definitions of App.~\ref{app: a} can be conveniently recast in the suggestive form
\begin{align}
    \tensor[^{(\phi)}]{\mathcal{E}}{_{\lambda\mu\nu}}+\tensor[^{(\chi_1)}]{\mathcal{E}}{_{\lambda\mu\nu}}+\tensor[^{(\chi_2)}]{\mathcal{E}}{_{\lambda\mu\nu}}+\tensor[^{(\nabla\phi)}]{\mathcal{E}}{_{\lambda\mu\nu}}+\tensor[^{(\nabla\chi_1)}]{\mathcal{E}}{_{\lambda\mu\nu}}\tensor[^{(\nabla\chi_2)}]{\mathcal{E}}{_{\lambda\mu\nu}}+\tensor[^{(\nabla T)}]{\mathcal{E}}{_{\lambda\mu\nu}}=0
    \label{eq: connection dec}
\end{align}
where we introduced the shortcut notation
\begin{align}
    \tensor[^{(\phi)}]{\mathcal{E}}{_{\lambda\mu\nu}} & = \phi \leri{ \frac{Q_\lambda g_{\mu\nu}-Q_\mu g_{\lambda\nu}}{2}-Q_{\lambda\mu\nu}+P_\mu g_{\lambda\nu}+t_{\nu\mu\lambda}}\\
    \tensor[^{(\chi_1)}]{\mathcal{E}}{_{\lambda\mu\nu}} & =\chi_1 \leri{ \frac{Q_\lambda T_{\mu\nu}-Q_\rho\tensor[]{T}{^\rho_\mu} g_{\lambda\nu}}{2}+g_{\nu\lambda}Q_{\rho\sigma\mu}T^{\rho\sigma}-Q_{\lambda\rho\mu}T\indices{_{\nu}^\rho}-Q_{\lambda\rho\nu}T\indices{_{\mu}^\rho}+\tensor[]{T}{^\rho_\mu}t_{\nu\rho\lambda}}\\
     \tensor[^{(\chi_2)}]{\mathcal{E}}{_{\lambda\mu\nu}} & = \chi_2 \leri{\frac{Q_\rho \tensor[]{T}{^\rho_\lambda} g_{\mu\nu} -Q_\mu\tensor[]{T}{_{\lambda\nu}}}{2}+\tensor[]{T}{^\rho_\lambda}t_{\nu\mu\rho}-Q_{\rho\mu\nu}\tensor[]{T}{^\rho_\lambda}+Q_{\mu\rho\nu}\tensor[]{T}{^\rho_\lambda}+P_\mu T_{\lambda\nu}}\\
     \tensor[^{(\nabla\phi)}]{\mathcal{E}}{_{\lambda\mu\nu}} & = - g_{\mu\nu}\nabla_\lambda \phi + g_{\lambda\nu} \nabla_\mu \phi \\
     \tensor[^{(\nabla\chi_1)}]{\mathcal{E}}{_{\lambda\mu\nu}} & = - T_{\mu\nu}\nabla_\lambda \chi_1 + g_{\lambda\nu} \tensor[]{T}{^\rho_\mu}\nabla_\rho \chi_1 \\
     \tensor[^{(\nabla\chi_2)}]{\mathcal{E}}{_{\lambda\mu\nu}} & = \leri{\tensor[]{\delta}{^\rho_\mu}T_{\lambda\nu}-g_{\mu\nu}\tensor[]{T}{^\rho_\lambda}}\nabla_\rho \chi_2\\
     \tensor[^{(\nabla T)}]{\mathcal{E}}{_{\lambda\mu\nu}} & = \chi_1 \leri{g_{\lambda\nu}\nabla_\rho \tensor[]{T}{^\rho_\mu}-\nabla_\lambda T_{\mu\nu}} -\chi_2 \leri{g_{\mu\nu}\nabla_\rho \tensor[]{T}{^\rho_\lambda}-\nabla_\mu T_{\lambda\nu}}\nonumber
\end{align}
This particular rearrangement of the equation for the connection is extremely useful when one is interested in perturbation theory on the Minkowski background (see Sec.~\ref{sec: 4}). In this peculiar case, indeed, we anticipate that the only terms to survive containing a non-minimal coupling between the geometry and the energy momentum tensor components are those ones carried by $\tensor[^{(\nabla T)}]{\mathcal{E}}{_{\lambda\mu\nu}}$ which, in agreement with what we discussed in Sec.~\ref{sec: 1}, are the only ones exhibiting differential operators acting on the source terms.
\\ \indent We now turn our attention to the variation of \eqref{eq: action scalartensor} with respect to the metric field. We remark that care must be paid to the fact that with respect to the purely metric case, the Riemann tensor and its traces are now solely defined in terms of the independent connection, so that their variation is identically zero. Moreover, the explicit appearance of the energy momentum tensor in the action implies that additional source terms are expected to arise carrying the non-minimal coupling with the geometry since, as it can be deduced from definition \eqref{eq: def Tmn}, we have in general that
\begin{equation}
    \frac{\delta T_{\mu\nu}}{\delta g^{\alpha\beta}}=-\frac{1}{2}\leri{g_{\mu\alpha}g_{\nu\beta}+g_{\mu\beta}g_{\nu\alpha}}\mat+g_{\mu\nu}\frac{\delta \mat}{\delta g^{\alpha\beta}}-2\frac{\delta^2\mat}{\delta g^{\mu\nu}\delta g^{\alpha\beta}}.
    \label{eq: var T mn}
\end{equation}
With this in mind, the variation of \eqref{eq: action scalartensor} with respect to the (inverse of) the metric can be organized as
\begin{align}
    \delta_g S=\frac{1}{2\kappa^2}\int d^4x\,\Bigl[&\sqrt{-g}\;\leri{\phi \,\delta_g R+\chi_1 \,\delta_g X+\chi_2 \,\delta_g Y+2\kappa^2 \delta_g\mathcal{L}_m}+\Bigr.\\
    &\Bigl.+\,\delta_g\g\leri{\phi R+\chi_1 X+\chi_2 Y-W+2\kappa^2 \mathcal{L}_m}\Bigr],
\end{align}
where each variation has the form
\begin{align}
    &\delta_g R=R_{(\mu\nu)}\delta g^{\mu\nu}\\
    &\delta_g X=\leri{-\mat R_{(\mu\nu)}+\frac{\delta\mat}{\delta g^{\mu\nu}}R-2\frac{\delta^2\mat}{\delta g^{\rho\sigma}g^{\mu\nu}}R^{(\rho\sigma)}+R_{(\mu\rho)}T\indices{_\nu^\rho}+R_{(\nu\rho)}T\indices{_\mu^\rho}}\delta g^{\mu\nu}\\
    &\delta_g Y= \leri{-\mat\tilde{R}_{(\mu\nu)}+\frac{\delta\mat}{\delta g^{\mu\nu}}R-2\frac{\delta^2\mat}{\delta g^{\rho\sigma}g^{\mu\nu}}\tilde{R}^{(\rho\sigma)}+T_{\rho\sigma}R\indices{^\rho_{(\mu}^\sigma_{\nu)}}+T\indices{_{\rho(\mu}}\tilde{R}\indices{^\rho_{\nu)}}}\delta g^{\mu\nu}\\
    &\delta_g \mat = \frac{\delta \mat}{\delta g^{\mu\nu}}\delta g^{\mu\nu}\\
    & \delta_g\g=-\frac{\g}{2}g_{\mu\nu}\delta g^{\mu\nu} \ .
\end{align}
Recall that, as discussed in Sec.~\ref{sec: 1}, we are denoting with $\tilde{R}_{\mu\nu}$ the co-Ricci curvature tensor. Putting all the variations together, we can eventually rewrite the equation for the metric in the compact form
\begin{align}
    \phi \,G_{(\mu\nu)} &+\chi_1 \Theta_{\mu\nu} +\chi_2 \tilde{\Theta}_{\mu\nu}-\frac{1}{2}g_{\mu\nu}\leri{\chi_1 X+\chi_2 Y-W}=\kappa^2 T_{\mu\nu}
    \label{eq: metric}
\end{align}
where we defined the terms encapsulating the geometry-matter coupling as
\begin{align}
    \Theta_{\mu\nu}&\equiv -\mat R_{(\mu\nu)}+\frac{\delta\mat}{\delta g^{\mu\nu}}R-2\frac{\delta^2\mat}{\delta g^{\rho\sigma}g^{\mu\nu}}R^{(\rho\sigma)}+T\indices{_\nu^\rho}R_{(\mu\rho)}+T\indices{_\mu^\rho}R_{(\nu\rho)}\\
    \tilde{\Theta}_{\mu\nu}  &\equiv -\mat\tilde{R}_{(\mu\nu)}+\frac{\delta\mat}{\delta g^{\mu\nu}}R-2\frac{\delta^2\mat}{\delta g^{\rho\sigma}g^{\mu\nu}}\tilde{R}^{(\rho\sigma)}+T_{\rho\sigma}\tilde{R}\indices{^\rho_{(\mu}^\sigma_{\nu)}}+T\indices{_{\rho(\mu}}\tilde{R}\indices{^\rho_{\nu)}}.
\end{align}
For the sake of future convenience we report also the trace of \eqref{eq: metric}, which reads as
\begin{align}
    -\lerisq{\phi +(\chi_1+\chi_2)\leri{\mat -\frac{\delta\mat}{\delta g^{\mu\nu}}g^{\mu\nu}}}R-2g^{\mu\nu}\frac{\delta^2\mat}{\delta g^{\rho\sigma}g^{\mu\nu}}\leri{\chi_1R^{(\rho\sigma)}+\chi_2\tilde{R}^{(\rho\sigma)}}+2W=\kappa^2 T.
    \label{eq: trace metric}
\end{align}
The equations for the scalar field are quite immediate to obtain and they just result in
\begin{equation}
    R=\frac{\partial W}{\partial \phi},\quad X= \frac{\partial W}{\partial \chi_1},\quad Y=\frac{\partial W}{\partial \chi_2},
    \label{eq: scalar fields}
\end{equation}
which, as discussed in Sec.~\ref{sec: 4}, will allow us to get rid at linear level of the spurious scalar degrees of freedom and to express the dynamics in a more concise form once combined with \eqref{eq: trace metric}. A few comments are now in order. As observed in the introduction, in the presence of a non-minimal coupling between geometry and matter the covariant conservation of the energy momentum tensor is not in general guaranteed, both for the metric and the Palatini case. That usually results in the violation of the equivalence principle, as it is induced by the appearance of four-forces in the geodesic equation, which spoil the feasibility of local inertial frames. We expect the same to occur also in our model, even if the na\"{i}ve evaluation of the (metric) divergence of \eqref{eq: metric} is not sufficient to specify the exact amount of the violation. This is due to the fact that tensors $G_{\mu\nu},\,\Theta_{\mu\nu}$ and $\tilde{\Theta}_{\mu\nu}$ are genuine metric-affine quantities built out of the Riemann curvature, whose components are expected to depend in turn on the energy momentum tensor by virtue of \eqref{eq: connection dec}. This implies that before obtaining the correct expression for $\tensor[^{(L)}]{\nabla}{_{\rho}}T^{\rho\mu}$ from \eqref{eq: metric}, it is necessary to solve \eqref{eq: connection dec} for the different components of torsion and non-metricity, which is in general expected to be quite a challenging task due to the non-trivial coupling between the metric-affine geometry and the energy momentum tensor. Moreover, as discussed in Sec.~\ref{sec: 5}, the presence of four-forces in the geodesic equation can be related to additional contributions in the Einstein-Vlasov equation describing the evolution of the distribution function of the medium, potentially leading to mathematical inconsistencies in the analysis. These issues will be addressed in the next sections, where we shall show how, at the first order in perturbation theory over a Minkowski background, the connection can be formally solved in terms of the linearized derivatives of the energy momentum tensor, providing a compact description of the propagation of gravitational waves.

\section{The linearized theory}\label{sec: 4}
In this section we assume the metric to be described by the perturbative expansion $g_{\mu\nu}\simeq \eta_{\mu\nu}+h_{\mu\nu}$, with $\eta_{\mu\nu}$ denoting the Minkowski spacetime and $h_{\mu\nu}$ encapsulating the gravitational wave excitation. Accordingly, we set the energy momentum tensor $T_{\mu\nu}$ and the matter Lagrangian $\mat$, together with their derivatives, to be vanishing at the lowest order, so that $T_{\mu\nu}\simeq \delta T_{\mu\nu}$ (and similarly for $\mat$). This automatically implies that quantities like $X_{\mu\nu}$ or $Y$ are at least of second order in perturbation, helping us in deeply simplifying the equation for the metric and the scalar fields. Furthermore, in agreement with a Minkowskian description at the background level, we consider the connection, and so torsion and non-metricity, to be of first order in perturbation, so that in general we can write
\begin{equation}
    \Gamma\indices{^\rho_{\mu\nu}}\simeq \tensor[^{(L)}]{\delta\Gamma}{^\rho_{\mu\nu}}(h)+\tensor[]{\delta N}{^\rho_{\mu\nu}},
\end{equation}
where $\tensor[^{(L)}]{\delta\Gamma}{^\rho_{\mu\nu}}(h)$ is the Levi-Civita connection for the metric perturbation $h_{\mu\nu}$, and $\tensor[]{\delta N}{^\rho_{\mu\nu}}$ denotes the first order distorsion tensor (see App.~\ref{app: a}). Scalar fields are taken as
\begin{equation}
    \varphi_i=\Bar{\varphi}_i+\delta\varphi_i,
\end{equation}
with $i$ standing for the different $\phi,\chi_1,\chi_2$, and $\Bar{\varphi}_i$ describing the costant background value of the fields. Under these hypotheses, the zeroth order equations for the metric and the scalar fields lead to the conditions
\begin{align}
    W_B=\frac{\partial W_B}{\partial \phi}=\frac{\partial W_B}{\partial \chi_1}=\frac{\partial W_B}{\partial \chi_2}=0,
\end{align}
where the subscript $B$ stands for evaluation at the background values $\Bar{\varphi}_i$. This in turn implies that the linearized equation for the metric perturbation can be recast in the form
\begin{equation}
    \tensor[^{(L)}]{\delta G}{_{\mu\nu}}(h)+\delta A_{\mu\nu}=\frac{\kappa^2}{\Bar{\phi}}\delta T_{\mu\nu},
    \label{eq: lin metric}
\end{equation}
where $\tensor[^{(L)}]{\delta G}{_{\mu\nu}}(h)$ is the first order metric Einstein tensor and $\delta A_{\mu\nu}$ is the contribution from torsion and non-metricity \eqref{eq: A tensor affine}, which at linear level takes the form
\begin{equation}
    \delta A_{\mu\nu}=\frac{1}{2}\Biggl[\partial_\rho\leri{\delta N\indices{^\rho_{\mu\nu}}+\delta N\indices{^\rho_{\nu\mu}}}-\partial\indices{_{\nu}}\delta N\indices{^\rho_{\mu\rho}}-\partial\indices{_{\mu}}\delta N\indices{^\rho_{\nu\rho}}-\eta_{\mu\nu}\leri{\partial\indices{_\rho} \delta N\indices{^{\rho\sigma}_\sigma}-\partial_\sigma \delta N\indices{^{\rho\sigma}_{\rho}}}\Biggr].
    \label{eq: lin A tensor}
\end{equation}
Turning now our attention to the equation for the connection \eqref{eq: connection dec}, we see that at the first order the only non-vanishing terms are displayed by
\begin{align}
    \tensor[^{(\phi)}]{\delta\mathcal{E}}{_{\lambda\mu\nu}} & = \Bar{\phi} \leri{ \frac{\delta Q_\lambda \eta_{\mu\nu}-\delta Q_\mu \eta_{\lambda\nu}}{2}-\delta Q_{\lambda\mu\nu}+\delta P_\mu \eta_{\lambda\nu}+\delta t_{\nu\mu\lambda}}\\
    \tensor[^{(\nabla\phi)}]{\delta\mathcal{E}}{_{\lambda\mu\nu}} & = - \eta_{\mu\nu}\partial_\lambda \delta\phi + \eta_{\lambda\nu} \partial_\mu \delta\phi \\
    \tensor[^{(\nabla T)}]{\delta\mathcal{E}}{_{\lambda\mu\nu}} & = \Bar{\chi}_1 \leri{\eta_{\lambda\nu}\partial_\rho \tensor[]{\delta T}{^\rho_\mu}-\partial_\lambda \delta T_{\mu\nu}} - \Bar{\chi}_2 \leri{\eta_{\mu\nu}\partial_\rho \tensor[]{\delta T}{^\rho_\lambda}-\partial_\mu \delta T_{\lambda\nu}}
    \label{eq: lin dec connection}
\end{align}
From these relations it is possible to derive a set of equations for the different vector components of non-metricity by successive contractions of $\tensor[]{\delta\mathcal{E}}{_{\lambda\mu\nu}}$ with $\eta^{\mu\nu}$ and $\,\eta^{\nu\lambda}$, resulting respectively in
\begin{align}
    &\Bar{\phi}\leri{\frac{1}{2}\delta Q_\mu+\delta P_\mu}+\leri{\Bar{\chi}_1-3\Bar{\chi}_2}\partial_\rho\delta T\indices{^\rho_\mu}-\Bar{\chi}_1 \partial_\mu\delta T-3\partial_\mu\delta\phi=0
    \label{eq: lin vect conn 1}\\
    &\Bar{\phi}\leri{-\frac{3}{2}\delta Q_\mu+3\delta P_\mu}+\leri{3\Bar{\chi}_1-\Bar{\chi}_2}\partial_\rho\delta T\indices{^\rho_\mu}+\Bar{\chi}_2 \partial_\mu\delta T+3\partial_\mu\delta\phi=0,
    \label{eq: lin vect conn 2}
\end{align}
while the contraction with $\epsilon^{\lambda\mu\nu\sigma}$ simply results in the condition $\delta S_\mu=0$, implying that at the linear level torsion is purely tensorial. From \eqref{eq: lin vect conn 1} and \eqref{eq: lin vect conn 2} it is possible instead to derive the explicit form of the first and second trace of non-metricity, that is
\begin{align}
    &\delta Q_\mu = \frac{1}{\Bar{\phi}}\leri{4\partial_\mu\delta\phi+\frac{8\Bar{\chi}_2}{3}\partial_\rho T\indices{^\rho_\mu}+\frac{3\Bar{\chi}_1+\Bar{\chi}_2}{3}\partial_\mu\delta T}
    \label{eq: sol delta Q}\\
    &\delta P_\mu = \frac{1}{\Bar{\phi}}\leri{\partial_\mu\delta\phi-\frac{3\Bar{\chi}_1-5\Bar{\chi}_2}{3}\partial_\rho T\indices{^\rho_\mu}+\frac{3\Bar{\chi}_1-\Bar{\chi}_2}{6}\partial_\mu\delta T}
    \label{eq: sol delta P}.
\end{align}
Furthermore, by adding up \eqref{eq: lin vect conn 1} and \eqref{eq: lin vect conn 2} one can get the interesting relation
\begin{equation}
    \Bar{\phi}\leri{4\delta P_\mu-\delta Q_\mu}+\leri{\Bar{\chi}_1-\Bar{\chi}_2}\leri{4\partial_\rho \delta T\indices{^\rho_\mu}-\partial_\mu\delta T}=0,
\end{equation}
which for $\Bar{\chi}_1=\Bar{\chi}_2$ shows that the vector metric-affine structure assumes, at this perturbation order, a Weyl configuration characterized by
\begin{equation}
    \tensor[]{\delta Q}{_\mu}=4\tensor[]{\delta P}{_\mu}.
\end{equation}
Before addressing the solution for the remaining purely tensor component of torsion and non-metricity, it can be useful to discuss the dynamical properties of the scalar fields of our scalar-tensor representation. At  first order in perturbations, indeed, their equations can be rearranged by Taylor expanding the potential around the minimum in $\Bar{\varphi}_i$ as
\begin{align}
    &\frac{\partial^2 W_B}{\partial \phi^2}\delta\phi+\frac{\partial^2 W_B}{\partial \phi \partial \chi_1}\delta\chi_1+\frac{\partial^2 W_B}{\partial \phi \partial \chi_2}\delta\chi_2=-\frac{\kappa^2}{\Bar{\phi}}\delta T\\
    &\frac{\partial^2 W_B}{\partial \phi \partial\chi_1}\delta\phi+\frac{\partial^2 W_B}{\partial \chi_1^2}\delta\chi_1+\frac{\partial^2 W_B}{\partial \chi_1 \partial \chi_2}\delta\chi_2=0\\
    &\frac{\partial^2 W_B}{\partial \phi \partial\chi_2}\delta\phi+\frac{\partial^2 W_B}{\partial \chi_1 \partial\chi_2}\delta\chi_1+\frac{\partial^2 W_B}{\partial \chi_2^2}\delta\chi_2=0,
\end{align}
where we used \eqref{eq: trace metric} for substituting $\delta R$. We see that from the last two equations it is possible to solve for $\delta\chi_1,\,\delta\chi_2$, which can be expressed in the compact form
\begin{equation}
    \delta\chi_1=\frac{C_{\phi\chi_1}}{C_{\phi\phi}}\delta\phi,\;\delta\chi_2=\frac{C_{\phi\chi_2}}{C_{\phi\phi}}\delta\phi
\end{equation}
where $C_{ij}$ is the cofactor of the Hessian matrix for the potential, evaluated on the background. By combining these expressions with the first one, we can eventually rewrite the three scalar perturbations as
\begin{equation}
    \delta\phi=-\frac{\kappa^2 C_{\phi\phi}}{\Bar{\phi}\det H}\delta T,\;\;\;\delta\chi_1=-\frac{\kappa^2 C_{\phi\chi_1}}{\Bar{\phi}\det H}\delta T,\;\;\;\delta\chi_2=-\frac{\kappa^2 C_{\phi\chi_2}}{\Bar{\phi}\det H}\delta T,\qquad \det H\neq 0
    \label{eq: sol lin phi}
\end{equation}
where we used the definition of the determinant in terms of cofactors. Such a result highlights that, at linear level, the scalar fields are not true dynamical degrees of freedom carrying independent excitations, but they can be entirely expressed in terms of the trace of the perturbed energy momentum tensor, in analogy with the Palatini formulation of $f(R)$ gravity. This, in turn, indicates that the non-metricity vectors \eqref{eq: sol delta Q}-\eqref{eq: sol delta P} can be completely rewritten in terms of derivatives acting on $T_{\mu\nu}$, i.e.
\begin{align}
    &\delta Q_\mu = \frac{1}{\Bar{\phi}}\lerisq{\frac{8\Bar{\chi}_2}{3}u_\mu+\leri{\frac{3\Bar{\chi}_1+\Bar{\chi}_2}{3}-4\alpha}w_\mu}
    \label{eq: sol delta Q}\\
    &\delta P_\mu = \frac{1}{\Bar{\phi}}\lerisq{-\frac{3\Bar{\chi}_1-5\Bar{\chi}_2}{3}u_\mu+\leri{\frac{3\Bar{\chi}_1-\Bar{\chi}_2}{6}-\alpha}w_\mu}
    \label{eq: sol delta P},
\end{align}
where we introduced the shortcut notation $u_\mu\equiv \partial_\rho \delta T\indices{^\rho_\mu}$, $w_\mu\equiv\partial_\mu\delta T$ and $\alpha\equiv \frac{\kappa^2 C_{\phi\phi}}{\Bar{\phi}\det H}$. 
Now, plugging into the linearized \eqref{eq: connection dec} the solutions for $\delta Q_\mu,\,\delta P_\mu$ and $\delta S_\mu$, we obtain an equation for the purely tensor part of the connection, i.e.
\begin{align}
    \Bar{\phi}&\leri{\delta \Omega_{\lambda\mu\nu}-\delta q_{\nu\mu\lambda}}\label{eq: conn tensor part}  = \\&=\eta_{\mu\nu}\leri{-\frac{\Bar{\chi}_1+2\Bar{\chi}_2}{9}u_\lambda+\frac{5\Bar{\chi}_1+\Bar{\chi}_2}{18}w_\lambda}+\eta_{\lambda\nu}\leri{\frac{2\Bar{\chi}_1+\Bar{\chi}_2}{9}u_\mu-\frac{\Bar{\chi}_1+5\Bar{\chi}_2}{18}w_\mu}+\nonumber\\&+\eta_{\lambda\mu}\leri{\Bar{\chi}_1-\Bar{\chi}_2}\leri{\frac{4u_\nu-w_\nu}{18}}-\Bar{\chi}_1\partial_\lambda \delta T_{\mu\nu}+\Bar{\chi}_2\partial_\mu\delta T_{\lambda\nu}.\nonumber
\end{align}
The procedure necessary for extracting from \eqref{eq: conn tensor part} the explicit form for $\delta\Omega_{\lambda\mu\nu}$ and $\delta q_{\nu\mu\lambda}$ is quite involved and requires a bit of algebraic manipulation, that we illustrate below in detail. For this purpose, let us rewrite \eqref{eq: conn tensor part} in the compact form
\begin{equation}
    \delta\Omega_{\lambda\mu\nu}-\delta q_{\nu\mu\lambda}=A_{\lambda\mu\nu},
\end{equation}
where the rank-three tensor $A_{\lambda\mu\nu}$ contains all the contributions that we assume it is possible to solve in terms of the matter and the other dynamical fields. From the symmetry properties of torsion and non-metricity (see App.~\ref{app: a}), it is straightforward to obtain the following identities
\begin{equation}
    \delta \Omega\indices{_{(\lambda\mu)\nu}}=A\indices{_{(\lambda\mu)\nu}},\qquad \delta q\indices{_{[\nu\mu]\lambda}}=-A\indices{_{\lambda[\mu\nu]}},
    \label{eq: step 1}
\end{equation}
which inserted back into the original equation leads to the relation
\begin{equation}
    \delta \Omega\indices{_{[\lambda\mu]\nu}}-\delta q\indices{_{(\nu\mu)\lambda}}=A\indices{_{[\lambda\mu]\nu}}-A\indices{_{\lambda[\mu\nu]}},
    \label{eq: step 2}
\end{equation}
where we used the fact that a generic rank-$n$ tensor can be always decomposed in its symmetric and antisymmetric parts for any couple of indices, i.e.
\begin{equation}
    B_{\sigma_1\cdots \sigma_{n-2}\mu\nu}=B_{\sigma_1 \cdots \sigma_{n-2}(\mu\nu)}+B_{\sigma_1 \cdots \sigma_{n-2}[\mu\nu]}.
    \label{eq: dec general tensor}
\end{equation}
Now, subtracting from \eqref{eq: step 2} the equivalent expression with the position of the indices $\mu$ and $\nu$ reversed, it is easy to demonstrate that
\begin{equation}
    \delta\Omega_{[\nu\mu]\lambda}=A_{\lambda[\nu\mu]}+A_{[\nu|\lambda|\mu]},
    \label{eq: step 3}
\end{equation}
which up to a rearrangement of the indices and by using the property \eqref{eq: dec general tensor} can be combined with \eqref{eq: step 1} to get
\begin{equation}
    \delta\Omega_{\lambda\mu\nu}=A_{\lambda(\mu\nu)}+A_{\mu[\lambda\nu]}+A_{\nu[\lambda\mu]} \ .
    \label{eq: sol omega}
\end{equation}
This allows us to derive the expression for the tensor part of torsion directly from \eqref{eq: step 1}, i.e.
\begin{equation}
    \delta q_{\nu\mu\lambda}=-A_{\nu[\mu\lambda]}+A_{\mu[\lambda\nu]}-A_{\lambda[\mu\nu]}.
    \label{eq: sol q}
\end{equation}
Thus, by taking into account the definition of $A_{\lambda\mu\nu}$, we can write down the expression for the tensor part of non-metricity and torsion, which read explicitly as
\begin{align}
    &\delta \Omega_{\lambda\mu\nu}  =\frac{\Bar{\chi}_1-\Bar{\chi}_2}{18\Bar{\phi}}\Bigl(\eta_{\mu\nu}\leri{-2u_\lambda+5w_\lambda}+\eta_{\lambda\nu}\leri{4u_\mu-w_\mu}+\eta_{\lambda\mu}\leri{4u_\nu-w_\nu}-18\partial_\lambda\delta T_{\mu\nu}\Bigr)\label{eq: nonmetr tensor part}\\
    &\delta q_{\lambda\mu\nu}  = -\frac{\Bar{\chi}_2}{3\Bar{\phi}}\Bigl(\eta_{\lambda\nu}\leri{u_\mu-w_\mu}-\eta_{\lambda\mu}\leri{u_\nu-w_\nu}+3\leri{\partial_\mu\delta T_{\lambda\nu}-\partial_\nu\delta T_{\lambda\mu}}\Bigr),\label{eq: tors tensor part}
\end{align}
Eventually, putting together the results of Sec.~\ref{sec: 4}, we can derive the final expressions for torsion and non-metricity:
\begin{align}
    &\delta Q_{\lambda\mu\nu} =\frac{\eta_{\mu\nu}}{\Bar{\phi}}\leri{\frac{2\Bar{\chi}_2}{3}u_\lambda+\leri{\frac{3\Bar{\chi}_1-\Bar{\chi}_2}{6}-\alpha}w_\lambda}-\frac{\Bar{\chi}_1-\Bar{\chi}_2}{\Bar{\phi}}\partial_\lambda\delta T_{\mu\nu}\label{eq: final nonmetricity}\\
    &\delta t_{\lambda\mu\nu}=-\frac{\Bar{\chi}_2}{3\Bar{\phi}}\Bigl(\eta_{\lambda\nu}\leri{u_\mu-w_\mu}-\eta_{\lambda\mu}\leri{u_\nu-w_\nu}\Bigr)-\frac{\Bar{\chi}_2}{\Bar{\phi}}\leri{\partial_\mu\delta T_{\lambda\nu}-\partial_\nu\delta T_{\lambda\mu}}.\label{eq: final tors}
\end{align}
We observe that for $\Bar{\chi}_1=\Bar{\chi}_2$ the tensor part of non-metricity is vanishing, so that we fully recover the results we illustrated for the vector part, i.e. the setting of an exact Weyl geometry configuration characterized by
\begin{equation}
    \delta Q_{\lambda\mu\nu}=\delta P_\lambda \eta_{\mu\nu},\qquad \text{if}\;\; \Bar{\chi}_1=\Bar{\chi}_2.
\end{equation}
For $\Bar{\chi}_2=0$ instead, we note that it is the torsion tensor which completely disappears, so that in this case the disformal and the distorsion tensors coincide, i.e.
\begin{equation}
    \delta N\indices{^\rho_{\mu\nu}}=\delta D\indices{^\rho_{\mu\nu}},\qquad \text{if}\;\;\Bar{\chi}_2=0.
\end{equation}
We remark that even when the energy momentum conservation holds at the linear level, i.e. $u_\mu=0$, torsion and non-metricity are still non-trivial. In this case, however, a quick look at \eqref{eq: lin metric} reveals that in order $u_\mu=0$ to be actually consistent, the metric-affine contribution stemming out from the complete Einstein tensor must be divergence-free, i.e. $\partial_\mu\delta A^{\mu\nu}=0$. Once the explicit solutions for torsion and non-metricity have been implemented, this results in possible dynamical constraints over the energy momentum tensor or the parameters ruling the non-minimal coupling of the geometry with matter, and the consistency of the theory must be a posteriori checked. In order to address such an issue, we thereby need to evaluate the form of the distorsion tensor, which by virtue of \eqref{eq: christoffel contorsion disformal}, \eqref{eq: decomposition contorsion} and \eqref{eq: decomposition disformal}, eventually results in
\begin{align}
    \delta N_{\lambda\mu\nu}=&\frac{1}{2\Bar{\phi}}\leri{\frac{2\Bar{\chi}_2}{3}u_\nu+\leri{\frac{3\Bar{\chi}_1-\Bar{\chi}_2}{6}-\alpha}w_\nu}\eta_{\mu\lambda}-\frac{\Bar{\chi}_1-\Bar{\chi}_2}{2\Bar{\phi}}\partial_\nu\delta T_{\lambda\mu}\label{eq: sol distorsion}\\
    &-\frac{1}{2\Bar{\phi}}\leri{\frac{\Bar{\chi}_1+\Bar{\chi}_2}{2}-\alpha}\leri{\eta_{\mu\nu}w_\lambda-\eta_{\lambda\nu}w_\mu}+\frac{\Bar{\chi}_1+\Bar{\chi}_2}{2\Bar{\phi}}\leri{\partial_\lambda\delta T_{\mu\nu}-\partial_\mu\delta T_{\lambda\nu}}\nonumber.
\end{align}
This allows us to finally get the complete expression for the term $\delta A_{\mu\nu}$, i.e.
\begin{equation}
    \delta A_{\mu\nu}=\frac{\alpha}{\Bar{\phi}}\leri{\partial_\mu\partial_\nu-\eta_{\mu\nu}\Box}\partial_\rho w^\rho+\frac{\Bar{\chi}_1+\Bar{\chi}_2}{2\Bar{\phi}}\leri{\Box\delta T_{\mu\nu}+\eta_{\mu\nu}\partial_\rho u^\rho}-\frac{\Bar{\chi}_1+2\Bar{\chi}_2}{4\Bar{\phi}}\partial_{(\mu} u_{\nu)}.
\end{equation}
We immediately see that by evaluating the divergence of the expression above and taking into account the definition of the vector $u^\mu$ and $w^\mu$ in terms of the derivatives of the energy momentum tensor, we arrive at
\begin{equation}
    \partial_\rho \delta A\indices{^\rho_\mu}=\frac{\Bar{\chi}_1}{4}\leri{\Box\partial_\rho \delta T\indices{^\rho_\mu}+\partial_\mu \partial_\rho\partial_\sigma\delta T^{\rho\sigma}}=\kappa^2\partial_\rho \delta T\indices{^\rho_\mu},
\end{equation}
which is identically vanishing for $\Bar{\chi}_1=0$. As discussed in Sec.~\ref{sec: 2} and Sec.~\ref{sec: 5}, the possible violation of the energy momentum tensor conservation ultimately leads to the appearance of additional four-forces in the Einstein-Vlasov equation, as a result of spoiling the geodesic motion. The explicit dependence of such forces on the momenta of the particles, which constitute the medium and interact with the gravitational wave, must be in general introduced by suitable hypotheses which could guarantee the closeness of the dynamical system. We postpone this peculiar case for a future investigation and in this work we address the case where energy momentum conservation is retained at  first order in perturbations, so that we set $\Bar{\chi}_1=0$. In this scenario, the non-minimal coupling of the geometry with the matter is entirely embodied, at lowest order, in the contraction of the co-Ricci tensor with the energy momentum tensor, and the metric-affine structure is displayed by
\begin{align}
    &\delta Q_{\lambda\mu\nu} =-\frac{\eta_{\mu\nu}}{\Bar{\phi}}\leri{\frac{\Bar{\chi}_2}{6}+\alpha}\partial_\lambda \delta T+\frac{\Bar{\chi}_2}{\Bar{\phi}}\partial_\lambda\delta T_{\mu\nu}\label{eq: final nonmetricity cons}\\
    &\delta t_{\lambda\mu\nu}=\frac{\Bar{\chi}_2}{3\Bar{\phi}}\Bigl(\eta_{\lambda\nu}\partial_\mu \delta T-\eta_{\lambda\mu}\partial_\nu\delta T\Bigr)-\frac{\Bar{\chi}_2}{\Bar{\phi}}\leri{\partial_\mu\delta T_{\lambda\nu}-\partial_\nu\delta T_{\lambda\mu}}.\label{eq: final tors cons}
\end{align}
Under these assumptions, it is possible to rearrange the linear equation for the metric perturbation in the following effective form 
\begin{equation}
    \tensor[^{(L)}]{\delta G}{_{\mu\nu}}(h)-\leri{\partial_\mu\partial_\nu-\eta_{\mu\nu}\Box}\frac{\delta\phi}{\Bar{\phi}}=\frac{\kappa^2}{\Bar{\phi}}\delta T_{\mu\nu}- \frac{\Bar{\chi}_2}{2\Bar{\phi}}\Box\delta T_{\mu\nu},
    \label{eq: lin metric eff}
\end{equation}
where we resort the solution of $\delta\phi$ discussed in Sec.~\ref{sec: 4}. The above expression makes it clear that the contribution of the field $\phi$ to the equations for the metric perturbations has the same form as one obtains in standard scalar-tensor theories, of which metric and Palatini $f(R)$ theories are examples. The novel contribution is then carried by the parameter $\Bar{\chi}_2$, which modifies the source term by introducing an additional d'Alembert operator acting on the energy momentum tensor, as foreshadowed in Sec.~\ref{sec: 1}. In order to extract from \eqref{eq: lin metric eff} the equation for the tensor modes of the gravitational spectrum, which in our dynamical framework are the only polarizations which can escape the medium and freely propagate in vacuum, it is useful to define the generalized trace reversed tensor
\begin{equation}
    \bar{h}_{\mu\nu}\equiv h_{\mu\nu}-\frac{1}{2}\eta_{\mu\nu}\leri{h+\frac{\delta\phi}{\Bar{\phi}}},
    \label{eq: tt gauge}
\end{equation}
with $h=\eta^{\mu\nu}h_{\mu\nu}$, which allows us to rearrange \eqref{eq: lin metric eff} in the final form
\begin{equation}
    \Box \bar h_{\mu\nu}=-2\tilde{\kappa}^2\delta T_{\mu\nu}+\tilde{\chi}\Box\delta T_{\mu\nu} \ .
    \label{eq: metric perturbation linearized tt}
\end{equation}
Note that here we imposed the gauge conditions $\partial_\mu\Bar{h}^{\mu\nu}=0=\Bar{h}$ and defined the effective gravitational constant $\tilde{\kappa}^2\equiv \frac{\kappa^2}{\Bar{\phi}}$ and the parameter $\tilde{\chi}\equiv\frac{\Bar{\chi}_2}{\bar{\phi}}$, encoding the effect of the non-minimal coupling of geometry with matter. In the following we will assume that $\bar{\phi}>0$, to guarantee that $\tilde{\kappa}^2$ is positive definite and gravity attractive. This, in turn, implies that the sign of $\bar{\chi}$ is the same as that of $\bar{\chi}_2$, allowing for a direct comparison of the results in terms of deviations from GR (see the discussion in Sec.~\ref{sec: 6}).

\section{The Einstein-Vlasov system}\label{sec: 5}
Equation \eqref{eq: metric perturbation linearized tt} represents the starting point of our analysis about the propagation of gravitational waves in media filled with matter, which we assume to be composed by relativistic neutral particles of mass $m$, described in the context of the kinetic theory by the probability distribution function $f(\vec{x},\vec{p},t)$. This is defined in the single-particle phase space\footnote{We adopt as canonical coordinates the contravariant components of the position vector $x^i$ and the covariant components of the momentum $p_i$ (See \cite{Weinberg:2003ur,Flauger:2017ged}).} and normalized in order to return the total number of particles $N$ when integrated over its entire domain: the number $dN$ of particles with positions between $\vec{x}$ and $\vec{x}+d\vec{x}$ and momenta between $\vec{p}$ and $\vec{p}+d\vec{p}$ is given by $dN=f(\vec{x},\vec{p},t)d\vec{x}\, d\vec{p}$. 
Before the interaction with the gravitational perturbation, the medium is supposed to have reached an equilibrium state, with some distribution $f_0$ and temperature $\Theta$. We emphasize that our investigation is centered on gravitational radiation with wavelengths much smaller than the length scale of variation of the thermodynamic properties of the medium, like density, pressure and temperature. Under this assumption, the background configuration of the particles can be reasonably well described by a homogeneous and isotropic distribution function. Then, in order to include relativistic effects for the massive particles of the medium, we fix the equilibrium configuration as a J\"{u}ttner-Maxwell distribution
\begin{equation}
    f_0(p)=\dfrac{n}{4 \pi m^2 \Theta K_2 \leri{x}}e^{-\frac{\sqrt{m^2+p^2}}{\Theta}},
    \label{max jutt}
\end{equation}
where we introduced the density of particles $n$, the modulus of the flat three momentum $p=\sqrt{\delta^{ij}p_i p_j}$, and the modified Bessel function of the second kind $K_\nu\leri{\cdot}$ of real index $\nu$, evaluated in $x\equiv \frac{m}{\Theta}$.
The evolution of the medium is encoded in the Vlasov equation for the distribution function $f(\vec{x},\vec{p},t)$, which reads as
\begin{equation}
  \dfrac{D f}{dt}=\dfrac{\partial f}{\partial t}+\dfrac{d x^m}{dt}\dfrac{\partial f}{\partial x^m}+\dfrac{dp_m}{dt}\dfrac{\partial f}{\partial p_m}=0,
\end{equation}
the latter being nothing more than Boltzmann equation in which the collision integral on the right-hand side is neglected. The hypothesis of a collisionless medium can be implemented when the mean free path of the particles is much greater than the total size of the system or, alternatively, when the rate of collisions is much smaller than $\frac{1}{\Delta T}$, being $\Delta T$ the global time-scale of observation, i.e. the total time of interaction between the gravitational waves and the medium.
Then, by means of the identity $\frac{d x^i}{dt}=\frac{p^i}{p^0}$, with $p^0=\sqrt{m^2+g^{ij}p_ip_j}$ representing the energy of the particle, the geodesic equation can be rewritten as
\begin{equation}
    \frac{dp^m}{dt}+\Gamma\indices{^m_{\alpha\beta}}\frac{p^\alpha p^\beta}{p^0}=\frac{m^2}{p^0}F^m,
\end{equation}
where $F^m$ represents the contribution of possible external forces, which cause deviations of the motion from being purely geodesic. These terms are typical of theories with a non-minimal coupling of geometry with matter, where they originate from the non conservation of the energy momentum tensor \cite{Harko:2010mv,Harko:2011kv,Barrientos:2018cnx,Fox:2018gop}. Even if as discussed in Sec.~\ref{sec: 4} we will eventually disregard these contributions, for the moment we keep them for the sake of generality, in order to elucidate their role in the mathematical well-posedness of the kinetic theory formulation. \\The time derivative of the covariant momentum $p_m$ can be rewritten as
\begin{equation}
    \dfrac{dp_m}{dt}=\frac{\partial g_{mn}}{\partial t}p^n+\frac{\partial g_{mn}}{\partial x^k}\frac{p^k p^n}{p^0}-g_{mn}\Gamma\indices{^n_{\alpha\beta}}\frac{p^\alpha p^\beta}{p^0}+\frac{m^2}{p^0}F_m,
\end{equation}
which taking into account the perturbative expansions for the metric and the four-force $F^m\simeq \delta F^m$ reduces to
\begin{equation}
    \dfrac{dp_m}{dt}=\dfrac{p^ip^j}{2p^0}\dfrac{\partial g_{ij}}{\partial x^m}+\frac{m^2}{p^0}F_m=\dfrac{1}{2p^0}\leri{p_ip_j\dfrac{\partial\bar{h}_{ij}}{\partial x^m}+\frac{m^2}{\Bar{\phi}}\dfrac{\partial \delta\phi}{\partial x^m}}+\frac{m^2}{p^0}\delta F_m \ .
\end{equation}
Here we simply lowered the indices of $p^i$ with $\eta_{ij}$, as $\frac{dp_m}{dt}$ is already of first order in perturbation, and we rewrote $h_{\mu\nu}$ as illustrated by \eqref{eq: tt gauge}. Initially, the distribution function is assumed to be some isotropic equilibrium solution $f_0\leri{p}$ of the unperturbed equation. Therefore, at $t=0$ we simply have $f(\vec{x},\vec{p},0)=f_0( \sqrt{g^{ij}(\vec{x},0)p_ip_j})$, which at the first order in perturbation results in 
\begin{equation}
   f(\vec{x},\vec{p},0)=f_0 \leri{p}-\dfrac{f_0'(p)}{2}\dfrac{p_ip_j}{p}h_{ij}(\vec{x},0),
\end{equation}
where $f_0'(p) \equiv \frac{d f_0}{dp}$ and we used the fact that at the first order $p_ip_j h^{ij}=p_ip_j h_{ij}$.
\\For $t>0$, the distribution function is perturbed by the gravitational wave, i.e.
\begin{equation}
   f(\vec{x},\vec{p},t)=f_0 \leri{p}-\dfrac{f_0'(p)}{2}\dfrac{p_ip_j}{p}h_{ij}(\vec{x},t)+\delta f (\vec{x},\vec{p},t),
\end{equation}
where $\delta f (\vec{x},\vec{p},t)$ is small with respect to the equilibrium configuration, that is $\frac{\delta f}{f_0}=\mathcal{O}(h)$. The linearized Vlasov equation for the perturbation $\delta f (\vec{x},\vec{p},t)$ is then given by:
\begin{equation}
    \frac{\partial \delta f }{\partial t}+\frac{p^m}{p^0}\frac{\partial \delta f}{\partial x^m}-\frac{f_0'(p) }{2p} \leri{p_ip_j\frac{\partial \bar{h}_{ij}}{\partial t}-\frac{p^2}{\Bar{\phi}}  \frac{\partial \delta \phi}{\partial t}-\frac{p^0 p^m }{\Bar{\phi}} \frac{\partial\delta \phi}{\partial x^m}}+\frac{m^2}{p^0}p^m\delta F_m=0.
  \label{eq: linear vlasov}
\end{equation}
and in order for the dynamical problem to be self-consistent, it must be accompanied by the equation for the metric \eqref{eq: metric perturbation linearized tt}, where the sources have been conveniently rewritten in terms of the distribution function $f(\vec{x},\vec{p},t)$. That can be done by means of the energy momentum tensor of a Vlasov gas, i.e.
\begin{equation}
    T_{ij}(\vec{x},t)=\frac{1}{\sqrt{-g}}\int d^3p\, \dfrac{p_ip_j}{p^0}f (\vec{x},\vec{p},t),
\end{equation}
which at first order on a Minkowski background is simply given by
\begin{equation}
T_{ij}(\vec{x},t)=\int d^3p\, \dfrac{p_ip_j}{p^0}\delta f (\vec{x},\vec{p},t),
\label{eq: def t mn minkowski}
\end{equation}
where\footnote{In the following we will always use the notation $d^3p$ instead of $d\vec{p}$, previously introduced.} $d^3p=dp_1dp_2dp_3$. The set of equations formed by \eqref{eq: metric perturbation linearized tt} and \eqref{eq: linear vlasov} represents a system of coupled differential equations for $\bar h_{ij}(t,z)$ and $f(\vec{x},\vec{p},t)$, which can be conveniently turned into an algebraic problem by performing a combined Fourier-Laplace transform on spatial and temporal coordinates, respectively. Furthermore, let us assume that the gravitational perturbation is travelling along the $z$ direction (and the perturbed distribution function accordingly), so that the indices $i,j$ take the value $x,y$, and the d'Alembert operator simply reduces to $\Box=-\partial_t^2+\partial_z^2$. Then, under these hypotheses it is possible to derive the perturbation $\delta f(\vec{x},\vec{p},t)$ from \eqref{eq: linear vlasov} by algebraic manipulation, resulting in
\begin{equation}
    \delta f^{(k,s)}(\vec{p})=\frac{\frac{f_0'(p)}{2 p}\leri{p_ip_j\leri{s\,\bar{h}_{ij}^{(k,s)}-\bar{h}_{ij}^{(k)}(0)}-\leri{p^2s+ikp_3p^0}\frac{\delta\phi^{(k,s)}}{\Bar{\phi}}+\frac{\delta\phi^{(k)}(0)}{\Bar{\phi}}p^2}}{s+ik \frac{p_3}{p^0}},
\label{eq: sol delta f in FL}
\end{equation}
with $\bar{h}_{ij}^{(k,s)}$ denoting the Fourier-Laplace component of the gravitational perturbation of wave number $k$ and Laplace complex variable $s=-i\omega$, where $\omega$ is the frequency. When the label $s$ is omitted, the simple spatial Fourier transform at time $t=0$ is considered, which we refer to as initial conditions and where we use the fact that the Laplace transform of the time derivative of order $n$ of a function $g(t)$ is given by
\begin{equation}
    s^n g^{(s)} - \sum_{l=1}^{n}s^{n-l}\partial_t^{l-1}g(0^-).
\end{equation}
In \eqref{eq: sol delta f in FL}, we omitted the term $\delta F^m$ because it must be related to the other dynamical quantities $\bar{h}_{ij}^{(k,s)}$, $\delta f^{(k,s)}(\vec{p})$, or $p^m$  in order to have a well-posed and not under-determined mathematical problem. Establishing the explicit form of $\delta F^m$ evades the purpose of this work, and in the following we simply assume that such four-forces are vanishing, which as discussed at the end of Sec.~\ref{sec: 4} is  consistent with the conservation of the energy momentum tensor in the linearized theory. Now, by rewriting the metric perturbation in the chosen gauge as
\begin{equation}\label{metric perturbation tt gauge}
 \bar{h}_{ij} = \begin{pmatrix}
 h_+ & h_\times & 0 \\
 h_\times & -h_+ & 0 \\
 0 & 0 & 0
\end{pmatrix},
\end{equation}
we see that \eqref{eq: sol delta f in FL} takes the explicit form
\begin{equation}
    \delta f^{(k,s)}(\vec{p})=\frac{f_0'(p)\leri{\leri{s\,\bar{h}_+^{(k,s)}-\bar{h}_+^{(k)}(0)}\leri{p_1^2-p_2^2}+2\leri{s\,\bar{h}_\times^{(k,s)}-\bar{h}_\times^{(k)}(0)}p_1 p_2}}{2 p\leri{s+ik \frac{p_3}{p^0}}},\label{eq: delta f FL sol}
\end{equation}
where we neglected the contribution due to the scalar perturbation $\delta\phi$. This is due to the fact that, as can be appreciated from \eqref{eq: sol delta f in FL}, terms proportional to the scalar field always come with $p^2$ or $p_3$ factors, implying that their integral in the space of momenta of \eqref{eq: def t mn minkowski} is odd and thus vanishing for symmetry reasons. We can therefore evaluate the Fourier-Laplace transform of \eqref{eq: metric perturbation linearized tt}, obtaining
\begin{equation}
    \leri{s^2+k^2}\bar{h}_{ij}^{(k,s)}-s\bar{h}_{ij}^{(k)}(0)=\int d^3 p \frac{p_i p_j}{p^0}\lerisq{2\tilde{\kappa}^2\delta f^{(k,s)}+\tilde{\chi}\leri{\leri{s^2+k^2}\delta f^{(k,s)}-\partial_t \delta f^{(k)}(0)}},
    \label{eq: vlasov h step 1}
\end{equation}
where we set to zero the initial conditions $\partial_t\bar{h}_{ij}^{(k)}(0)$ and $\delta f^{(k)}(0)$. We can then appreciate that even if, in principle, $\delta f^{(k,s)}$ depends on both the polarization states $h_+,\;h_\times$, the equations for the different gravitational modes are actually decoupled because off-diagonal terms are identically vanishing again for symmetry reasons once integration is performed in the momenta $p_1, p_2, p_3$. Furthermore, we add that the last initial condition $\partial_t \delta f^{(k)}(0)$ can be analogously set to zero, in that as determined from the Vlasov equation in the Fourier space, we have
\begin{equation}\label{derzero}
    \partial_t\delta f^{(k)}(0)=-\dfrac{f_0'(p)}{2p}\frac{ik p^0p_3}{\Bar{\phi}}\delta\phi^{(k)}(0),
\end{equation}
which as proved by \eqref{eq: sol lin phi} and \eqref{eq: def t mn minkowski} is vanishing once $\delta f^{(k)}(0)$ is set to zero. We can then solve \eqref{eq: vlasov h step 1} for the metric perturbation, i.e.
\begin{equation}
    \bar{h}_{ij}^{(k,s)}=\frac{s \bar{h}_{ij}^{(k)}(0) +\int d^3 p \frac{p_i p_j}{p^0}\leri{2\tilde{\kappa}^2+\tilde{\chi}\leri{s^2+k^2}}\delta f^{(k,s)}}{s^2+k^2},
    \label{eq: vlasov h step 2} 
\end{equation}
and casting eventually \eqref{eq: delta f FL sol} into \eqref{eq: vlasov h step 2} yields the self-consistent equation for the tensor perturbation within the medium, that is
\begin{equation}
    \bar{h}_{+,\times}^{(k,s)}=\dfrac{s-\frac{\pi}{4}\leri{2\tilde{\kappa}^2+\tilde{\chi}\leri{s^2+k^2}}\int dp_3 d\rho  \, \frac{\rho^5 f'_0(p)}{p}\frac{1}{p^0s+ikp_3}}{\leri{s^2+k^2}\epsilon(k,s)}\;\bar{h}_{+,\times}^{(k)}(0),
\end{equation}
where we introduced cylindrical coordinates in the momentum space, i.e.
\begin{equation}
    p_1=\rho \cos \varphi,\quad p_2=\rho \sin \varphi,\quad p_3=p_3,
\end{equation}
with integration performed in the volume $dV=d\varphi d\rho dp_3$, with
\begin{equation}
    \varphi\in[0,2\pi],\;\rho\in[0,+\infty),\;p_3\in(-\infty,+\infty).
\end{equation}
We defined the gravitational dielectric function as
\begin{equation}
    \epsilon(k,s)=1-\frac{\pi}{4}\leri{\frac{2\tilde{\kappa}^2}{s^2+k^2}+\tilde{\chi}}s\int dp_3 d\rho  \, \frac{\rho^5 f'_0(p)}{p}\frac{1}{p^0s+ikp_3}
\end{equation}
where the correction to the standard GR result led by the parameter $\tilde{\chi}$ is manifest (compare with the results of \cite{Moretti:2020kpp,Bombacigno:2022naf}).\\ 

\begin{figure}[h]
    \centering
   \begin{tikzpicture}
   \draw [red, dashed] (-5,0) to (-4,0);
   \draw [red] (-4,0) to (-3,0) node[anchor=north]{$\nexists \, v^2_p$};
   \draw [red] (-3,0) to (-2,0);
   \draw (-2,-0.25) to (-2,0.25) node[anchor=south]{$\tilde{\chi}=\chi_{deg}$};
   \draw [ultra thick] (-2,0) to (-1,0) node[anchor=north]{$v^2_p>1$};
   \draw [ultra thick] (-1,0) to (0,0);
   \draw (0,-0.25) to (0,0.25) node[anchor=south]{$\tilde{\chi}=0$};
   \draw [ultra thick] (0,0) to (1,0) node[anchor=north]{$v_p^2>1$};
   \draw [ultra thick] (1,0) to (2,0) ;
   \draw [dashed, ultra thick] 
   (2,0) edge[-latex] (3,0);
   \draw (3.1,0) to (3.1,0) node[right]{$\tilde{\chi}$};
  \end{tikzpicture}
    \caption{Propagation for the branch $\omega_{r,+}$ for $\forall\;k$}
    \label{fig: 1}
\end{figure}
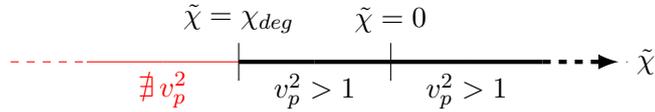

\begin{figure}[h]
    \centering
   \begin{tikzpicture}
   \draw [dashed, ultra thick] (-5,0) to (-4,0);
   \draw [ultra thick] (-4,0) to (-3,0) node[anchor=north]{$v^2_p<1$};
   \draw [ultra thick] (-3,0) to (-2,0);
   \draw [red] (-2,-0.25) to (-2,0.25) node[anchor=south]{$\tilde{\chi}=\chi_{deg}$};
   \draw [ultra thick] (-2,0) to (-1,0) node[anchor=north]{$v^2_p<1$};
   \draw [ultra thick] (-1,0) to (0,0);
   \draw (0,-0.25) to (0,0.25) node[anchor=south]{$\tilde{\chi}=0$};
   \draw [red] (0,0) to (1,0) node[anchor=north]{$\nexists\;v_p^2$};
   \draw [red] (1,0) to (2,0) ;
   \draw [dashed, red] (2,0) edge[-latex] (3,0);
    \draw (3.1,0) to (3.1,0) node[right]{$\tilde{\chi}$};
  \end{tikzpicture}
    \caption{Propagation for the branch $\omega_{r,-}$ for $k^2>k_0^2$}
    \label{fig: 2}
\end{figure}
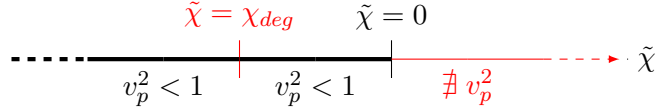

We now focus our investigation in the so-called \textit{weak damping scenario} \cite{Landau:1946jc,lifshitz1995physical}, characterized by a frequency with an imaginary part much smaller than the real part, i.e. $|\omega_i|\ll|\omega_r|$. Under this hypothesis, the oscillation period of  the gravitational wave is still much smaller than the typical damping time, thereby preventing the formation of transient signals which decay too much rapidly for being detected. In this setting, the dispersion relation $\omega_r=\omega_r(k)$ is derived directly from the real part of the dielectric function by solving
\begin{equation}
    \Re (\epsilon)(k,\omega_r)=0,
    \label{real part epsilon}
\end{equation}
which allows us to eventually determine the damping coefficient directly as
\begin{equation}
    \omega_i=-\left.\frac{ \Im(\epsilon_{L,R})}{\frac{\partial\Re (\epsilon_{L,R})}{\partial\omega}}\right|_{\omega=\omega_r}.
    \label{imaginarypart}
\end{equation}
For the J\"uttner-Maxwell background distribution \eqref{max jutt} the dielectric functions are given by:
\begin{align}
    \epsilon (k,\omega_r)=1-\frac{ n \leri{\frac{2\tilde{\kappa}^2}{k^2}+\tilde{\chi}\leri{1-v_p^2}}}{8 m^2\Theta^2K_2\leri{x}}\leri{\dfrac{v_p}{1-v_p^2}}^2\int d\rho dp_3 \, \dfrac{\rho^5e^{-\frac{\sqrt{m^2+\rho^2+p_3^2}}{T}}}{p_3^2-\frac{v_p^2 }{1-v_p^2}(m^2+\rho^2)}, \label{eq: dielectric jutt}
\end{align}
where $v_p\equiv\frac{\omega_r}{k}$ is the phase velocity. We observe that along the integration path of $p_3$ a pair of poles located at $p_3=\pm\sqrt{\frac{v_p^2}{1-v_p^2}(m^2+\rho^2)}$ can appear, whenever the condition $v_p<1$ holds. This implies the existence of an imaginary part for dielectric functions, which can then be evaluated by applying the residue theorem. The range of validity of the inequality $v_p<1$ can be established, however, only by solving \eqref{real part epsilon} for $v_p$, and it usually results in phenomenological constraints relating the parameters of the model and the physical properties of the medium (see discussion in \cite{Moretti:2020kpp,Bombacigno:2022naf}). In order to obtain $\omega_r$, we follow the standard approach of plasma physics \cite{lifshitz1995physical} by expanding the denominator of the integrals in \eqref{eq: dielectric jutt} up to the second order in $p_3$. This amounts to assuming  $\frac{p_3}{v_p}\sqrt{\frac{1-v_p^2}{m^2+\rho^2}}\ll 1$, which corresponds to having a phase velocity for the wave much greater than the thermal velocity of the medium. It has to be remarked that this assumption typically holds for material media in weak field regime, such as galactic and Solar System dark matter halos. In fact, in the presence of strong gravity scenarios, characterized by $x\simeq 1$ and corresponding to high density and temperature, this hypothesis is not well grounded and numerical techniques of integration of the dielectric function are usually required (see \cite{Moretti:2020kpp}). The real part of the dielectric function is then given by:
\begin{equation}
    \Re (\epsilon)=1+\frac{2\omega_0^2}{x^2}\leri{1-\frac{\tilde{\chi}\leri{\omega^2-k^2}}{2\tilde{\kappa}^2}}\leri{\frac{\gamma (x)}{\omega^2}-\frac{x}{\omega^2-k^2}}
    \label{eq: real dielectric}
\end{equation}
where we introduced the proper frequency of the medium $\omega_0^2=\tilde{\kappa}^2 n m$ (see \cite{Moretti:2020kpp}) and $\gamma(x)=\frac{K_1(x)}{K_2(x)}$. Then, by solving $\Re\leri{\epsilon}=0$ we can finally obtain the dispersion relation, i.e.
\begin{equation}
\omega^2_{r,\pm}(k)=k^2\leri{\frac{-B\pm\sqrt{B^2-4AC}}{2A}},
\label{eq: dispersion solution}
\end{equation}
where we defined the coefficients for the biquadratic equation in $\omega_r$ as
\begin{align}
    &A(x,k,\tilde{\chi})\equiv 1+\frac{\tilde{\chi}\omega_0^2}{\tilde{\kappa}^2}\frac{x-\gamma}{x^2} \\
    &B(x,k,\tilde{\chi})\equiv -1-\frac{\tilde{\chi}\omega_0^2}{\tilde{\kappa}^2}\leri{\leri{1+\frac{2\tilde{\kappa}^2}{\tilde{\chi}k^2}}\frac{x-\gamma}{x^2}-\frac{\gamma}{x^2}}\\
    &C(x,k,\tilde{\chi})\equiv -\frac{\tilde{\chi}\omega_0^2}{\tilde{\kappa}^2} \leri{1+\frac{2\tilde{\kappa}^2}{\tilde{\chi}k^2}}\frac{\gamma}{x^2}.
\end{align}
As previously discussed, the necessary condition for the propagation throughout the medium to be allowed and the kinematic damping to set is embodied in the constraint $0<v_p^2<1$. This relation implies that for each branch of \eqref{eq: dispersion solution}, different regions of the $k$-space are selected where wave propagation is respectively possible or forbidden and phase velocity sub- or super-luminal. These domains can be conveniently presented in terms of the value assumed by the parameter $\tilde{\chi}$ with respect to the critical point
\begin{equation}
    \chi_{deg}= -\frac{\tilde{\kappa}^2x^2}{(x-\gamma)\omega_0^2}<0,
\end{equation}
corresponding to the degenerate case $A(x,k,\chi_{deg})=0$ where \eqref{eq: real dielectric} boils down to a simple quadratic equation for $\omega_r$ (see the discussion in Subsec.~\ref{subsec: 5.1}). Then, the analysis reveals that the plus branch $\omega_{r,+}$ exists only for $\tilde{\chi}>\chi_{deg}$, with no restriction on the allowed wavenumbers $k$ and a phase velocity always superluminal, preventing the possibility of kinematic damping between the gravitational wave and the medium (Fig.~\ref{fig: 1}). This is consistent with the result of GR, which we remind to be recovered for $\tilde{\chi}=0$.
\\The minus branch $\omega_{r,-}$ is instead available only for $\tilde{\chi}<0$, provided $\tilde{\chi}\neq \chi_{deg}$, with a cut-off on the wavenumbers given by $k^2>k_0^2\equiv\frac{2\tilde{\kappa}^2}{|\tilde{\chi}|}$. Here the phase velocity is always subluminal, implying the feasibility of the Landau phenomenon for all the propagating frequencies (Fig.~\ref{fig: 2}).
\\It is worth noticing that in the respective limits $\tilde{\chi}\rightarrow + \infty$ and $\tilde{\chi}\rightarrow - \infty$, both branches $\omega_{r,+}$ and $\omega_{r,-}$ tend to the same value
\begin{equation}
    \omega_{r,\infty}^2=k^2,
\end{equation}
representing the linear non-dispersive wave propagation of GR in vacuum, where both the phase velocity and the group velocity $v_g=\frac{d\omega}{dk}$ are equal to the speed of light. We remark that such a limit configuration must be taken carefully, as a quick analysis of \eqref{eq: metric perturbation linearized tt} reveals that for the consistency of the perturbative expansion, i.e. $\tilde{\chi}\Box\delta T_{\mu\nu}$ still of order $\mathcal{O}(h)$, the matter perturbation has to vanish more rapidly than the growth of the parameter $\tilde{\chi}$, ultimately resulting in the GR equation for the tensor modes in vacuum $\Box h_{\mu\nu}=0$. 

A more interesting limit is represented by the case of a Newtonian medium, corresponding to $x\rightarrow + \infty$, where the two branches reduce to
\begin{align}
    &\omega_{r,+}^2 = k^2,\quad \omega_{r,-}^2= 0.
\end{align}
In this case, indeed, the critical value $\chi_{deg}$ is moved to $-\infty$, the cut-off wavenumber for the $\omega_{r,-}$ branch vanishes and the coefficients $A,B,C$ boil down to $1,-1,0$. We can calculate now the imaginary part of \eqref{eq: dielectric jutt}, by re-iteratively integrating by parts in the new variable $\rho=m \sqrt{y^2-1}$ and resorting to the residues theorem for the integration in $p_3$, leading to (see discussion in \cite{Moretti:2020kpp,Bombacigno:2022naf}):
\begin{equation}
    \Im (\epsilon)=-\frac{\pi n m v_p}{x^3 K_2(x)}\leri{\frac{2\tilde{\kappa}^2}{k^2}+\tilde{\chi}\leri{1-v_p^2}}\leri{x^2+3x\sqrt{1-v_p^2}+3\leri{1-v_p^2}}e^{-\frac{x}{\sqrt{1-v_p^2}}}
\end{equation}
Then, from \eqref{imaginarypart} with the real part given by \eqref{eq: real dielectric}, it is possible to derive the damping coefficient for the branch $\omega_{r,-}$, the only one where the Landau damping phenomenon is actually feasible:
\begin{equation}
    \omega_i=\frac{\pi k^3\leri{\tilde{\chi}\leri{1-v_p^2}+\frac{2\tilde{\kappa}^2}{k^2}} \leri{x^2+3x\sqrt{1-v_p^2}+3\leri{1-v_p^2}}}{4x\tilde{\kappa}^2K_2(x)\leri{\frac{x}{\leri{1-v_p^2}^2}-\frac{\gamma}{v_p^4}\leri{1+\frac{k^2 \tilde{\chi}}{2\tilde{\kappa}^2}}}}e^{-\frac{x}{\sqrt{1-v_p^2}}},
\end{equation}
which for $k^2>k_0^2$, where the dispersion relation $\omega_{r,-}$ exists, is identically negative, guaranteeing the absence of instabilities due to the enhancement of specific propagating wavenumber $k$.

\subsection{The degenerate case $A(x,k,\chi_{deg})=0$}\label{subsec: 5.1}
When $\tilde{\chi}$ assumes the critical values $\chi_{deg}$ the equation determining the real part of the frequency simply reduces to
\begin{equation}
    B \omega_r^2+Ck^2=0
\end{equation}
so that the dispersion relation in this case is described by a single branch given by
\begin{equation}
    \omega_r^2=\frac{k^2 \gamma \leri{x^2k^2-2\omega_0^2\leri{x-\gamma}}}{2\omega_0^2\leri{x-\gamma}^2+\gamma x^2k^2}.
\end{equation}
It is easy to show that $0<\omega_r^2<1$ for
\begin{equation}
    k^2>k_{0,deg}^2\equiv \frac{2\omega_0^2\leri{x-\gamma}}{x^2},\quad x>0,
\end{equation}
the latter being automatically satisfied by definition. This results in the damping coefficient
\begin{equation}\label{omegai}
    \omega_i=\frac{\pi k^3\leri{\frac{2}{ k^2}-\frac{x}{x-\gamma}\frac{1-v_p^2}{\omega_0^2}} \leri{x^2+3x\sqrt{1-v_p^2}+3\leri{1-v_p^2}}}{4x K_2(x)\leri{\frac{x}{\leri{1-v_p^2}^2}-\frac{\gamma}{v_p^4}\leri{1-\frac{k^2 x}{2\omega_0^2\leri{x-\gamma}}}}}e^{-\frac{x}{\sqrt{1-v_p^2}}}.
\end{equation}


\section{Estimates}\label{sec: 6}
The purpose of this section\footnote{Here we restore SI units.} is to provide quantitative estimates of the predicted damping and the theory parameters typical values for realistic scenarios. In particular, we are interested in assessing the effective feasibility of the regime in which collisionless damping is indeed possible, i.e. when the $\omega_-$ branch of the dispersion relation is accessible for $\tilde{\chi}<0$, $\tilde{\chi}\neq \chi_{deg}$ and $|k|>|k_0|$. We start the discussion by observing that the value of $k_0$ can be constrained by considering the Newtonian limit of \eqref{eq: lin metric eff}, with the observational bounds obtained for the Eddington-inspired-Born-Infeld gravity (see the analysis of Sec.~III of \cite{Avelino:2012qe}). It can be checked, indeed, that by adopting the gauge
\begin{equation}
    h\indices{^\mu_{\nu,\mu}}-\frac{1}{2}h\indices{^\mu_{\mu,\nu}}=\frac{\partial_\nu\delta\phi}{\phi_0},
\end{equation}
it is possible to derive the following modified Poisson equation for the Newtonian potential

\begin{equation}
    \triangle \Psi_N=4\pi G\rho-\leri{\frac{4\pi G C_{\phi\phi}}{\bar{\phi}\det H}+\frac{c^4\tilde{\chi}}{4}}\triangle \rho,
    \label{eq: mod poisson}
\end{equation}
where we defined $h_{00}\equiv  -\frac{2}{c^2}\Psi_N$, and we resorted to the solution of $\delta\phi$ in terms of the perturbed energy momentum tensor. We also introduced the static weak-field approximation, namely
\begin{equation}
    \Box\simeq \nabla^2\equiv\triangle,\quad \delta T_{00}\simeq c^2\rho \, ,\quad \delta T\simeq -c^2\rho\, ,\quad \delta T_{ij}\simeq 0.
\end{equation}
Then, a direct comparison of (3.5) in \cite{Avelino:2012qe} with \eqref{eq: mod poisson} allows us to write down a constraint for $k_0$ from
\begin{equation}
    \epsilon=-2\leri{\frac{C_{\phi\phi}}{\bar{\phi}^2\det H}-\frac{1}{\bar{\phi}k_0^2}}<6\times 10^{5}\; \text{m}^2,
\end{equation}
which under the conservative hypotheses of $\bar{\phi}\simeq 1$ and $\frac{C_{\phi\phi}}{\det H}\simeq 0$ ultimately results in
\begin{equation}
    k_0>k_0^{min}=1.7\times 10^{-3}\; \text{m}^{-1}.
\end{equation}
From the definition of $k_0$, it is therefore possible to put an upper bound on the parameter $\chi_2$, i.e.
\begin{equation}\label{boundchi}
|\chi_2|\in\leri{0,\chi_2^{max}},\quad\text{with}\quad \chi_2^{max}\equiv\frac{2\kappa^2}{(k_0^{min})^2}=1.4\times 10^{-37}\; \text{Pa}^{-1}.
\end{equation}
Let us now focus on the other relevant parameter introduced in our analysis, namely $\chi_{deg}$: we see from its definition that its value is determined by the physical properties of the material medium traversed by gravitational radiation. Recalling that the dispersion relation and the damping rate have been evaluated under the hypothesis of large $x$, i.e. the medium is assumed to be scarcely relativistic, we can safely approximate $\chi_{deg}$ as
\begin{equation}
    |\chi_{deg}|\approx \frac{x}{c^2n\,m}.
\end{equation}
By considering the present-day value of the dark matter density on cosmological scales, amounting to an approximate value $\rho=nm\approx 10^{-27}\, \frac{\text{kg}}{\text{m}^3}$, and a large $x=100$ value, coherent with a  ``cold" model for dark matter, we calculate
\begin{equation}
    |\chi_{deg}|\approx 10^{12} \text{Pa}^{-1}.
    \end{equation}
Hence, it is straightforward to deduce that the degenerate case $\chi=\chi_{deg}$ is not actually achievable when the propagation of gravitational waves in interaction with dark matter in the late Universe is considered. An ulterior possibility is to apply our findings to the case of gravitational radiation propagating in dark matter on cosmological time-scales. Indeed, under this assumption many quantities involved in our formulae are actually dependent on the cosmological redshift $z$, namely 
\begin{align}
    k&=k^{(0)} (1+z) \\
         \omega_0&=\omega_0^{(0)} (1+z)^{\frac{3}{2}}\\
         x&=x^{(0)}(1+z)^{-1},
\end{align}
where present-day values are indicated by the superscript $(0)$. We set $k^{(0)} \approx 10^{-6}\, \text{m}^{-1}$ in order to deal with radiation which can be detected with ground-based interferometers, while we keep $x^{(0)}=100$ and $\rho^{(0)}=10^{-27}\, \frac{\text{kg}}{\text{m}^3}$. Plugging these values into the damping rate \eqref{omegai} yields an expression for $\omega_i$ which solely depends on the redshift $z$. We fix $|\tilde{\chi}|=10^{-37}$, a value compatible with the bound \eqref{boundchi} under the conservative assumption $\bar{\phi}\approx 1$. In this scenario we calculate a maximum damping rate $\omega_i \approx -10^{15}\, \text{Hz}$, a huge value which would be connected to a total absorption phenomenon. However, the redshift at which the interaction takes place is approximately $z\approx 10^{15}$, where $x\ll 1 $. In this regime, in which the relativistic effects are heavily impacting, the hypotheses on which we base our analysis are not well grounded and the formulae we derived lose their predictivity. Moreover, we find that the interaction time is uniquely determined by the value of $|\tilde{\chi}|$ and by choosing a value smaller than $10^{-37}$ would imply an interaction located at even larger $z$. Hence, given the constraint on $|\tilde{\chi}|$ displayed in \eqref{boundchi}, we claim that no damping effect is expected on gravitational waves with frequencies around $100 \, \text{Hz}$, at least when the interaction with cosmological dark matter is considered.   

\section{Summary and discussion of the results}\label{sec: 7}
In this work we investigated the propagation of gravitational waves in matter, when a non-minimal coupling between geometry and matter  is present, and a generic metric-affine spacetime is considered. In particular, we focused our analysis on a particular class of theories, where the coupling is entirely carried by the contractions of the energy momentum tensor with the Ricci and the co-Ricci curvature. 

Adopting a Palatini perspective, i.e. considering a vanishing hypermomentum, we solved exactly the linearized equation for the connection, showing how it depends on the derivatives of the energy momentum tensor. In this respect, we extended the previous results of the literature \cite{Fox:2018gop,Olmo:2019flu}, allowing for the presence of a non-trivial torsion, which turned to be characterized by a purely tensor component. Notably, during the process we developed an algorithmic procedure capable of extracting the tensor part of torsion and non-metricity, which has the merit of being applicable whenever a linear algebraic equation for the connection is practicable and the source terms can be clearly separated. We observe that the resulting equation for the connection can be also viewed in terms of an effective hypermomentum tensor, encapsulating the contribution from the energy momentum tensor and its derivatives. This allows in principle for a possible identification of the effective spin, shear and dilation components, by analogy with what is done in modified gravity for FRLW cosmology, where an effective energy density and pressure are usually introduced.

An interesting outcome of our analysis is the emerging condition assuring the conservation of the energy momentum tensor at linear level. Notably, this turns out to depend on the coupling with the Ricci tensor, being instead insensitive to the parameter associated with the co-Ricci curvature. Such a result implies that the plain conversion of $f(R,R_{\mu\nu}T^{\mu\nu})$ gravity into the Palatini formalism does not seize the full complexity of the geometry-matter couplings available in metric-affine scenarios. In particular, it is worth noticing that torsion is completely vanishing only when the co-Ricci tensor is neglected, in agreement with \cite{Fox:2018gop,Olmo:2019flu}, where torsion is set to zero from the beginning. These properties, together with the possibility to rewrite the scalar fields of the Jordan frame in terms of the energy momentum tensor trace, guarantee that no additional degrees of freedom are propagated at the linear level, in analogy to Palatini $f(R)$ gravity.
Even if a detailed discussion of the non-linear theory evades the purposes of the current work, it is still possible to infer some general properties we expect to hold, when the full non-minimal coupling between matter and geometry is taken into account. As it can be appreciated from \eqref{eq: connection dec}, in this regime the equation for the connection is algebraic and linear in the torsion and non-metricity components, with the energy momentum tensor and the scalar fields of the Jordan representation only appearing in the form of linear couplings or through their first covariant derivatives. That implies that the effective equations for the metric and the scalar fields can contain at most second order derivatives for the fields $\phi,\chi_1,\chi_2$ and third order derivatives for the matter fields described by $T_{\mu\nu}$. This clearly raises the problem of possible dynamical instabilities related to the matter degrees of freedom, as it occurs in the corresponding metric formulation \cite{Ayuso:2014jda}, and constitutes material for further investigations.\\
Insofar the linear theory is considered, the main effect of the additional d'Alembert operator on the energy momentum tensor is the emergence of two distinguished branches, that we called $\omega_{\pm}(k)$, for the dispersion relations describing the propagation of gravitational waves in the medium. The existence of each branch is determined by the value of the parameter $\tilde{\chi}$, related to the strength of the coupling between the co-Ricci curvature and the energy momentum tensor, with respect to the critical values $\tilde{\chi}=0$ and $\chi_{deg}$. In particular, we found that for $\tilde{\chi}>0$ the only viable branch is $\omega_+$, with the propagation characterized by a superluminal phase velocity and no restriction on the allowed wave-numbers $k$. For $\chi_{deg}<\tilde{\chi}<0$ both the branches become available, but they display different properties. While the $\omega_+$ channel retains the same characteristics as the previous case, the branch $\omega_-$ only allows propagation of modes with $k^2>k_0^2$, where phase velocity is subluminal and Landau damping can occur. In particular, the cut-off scale $k_0$ is set by the inverse of the parameter $\tilde{\chi}$, so that for $\tilde{\chi}\to 0^-$ the value of $k_0$ diverges and $\omega_-$ disappears. Finally, for $\tilde{\chi}<\chi_{deg}$ we solely have the $\omega_-$ branch, where the phase velocity is still subluminal and with the same lower bound on the wave-numbers $k$. In this case, the limit $\tilde{\chi}\to -\infty$ corresponds to $k_0 \to 0$, with no restrictions on the propagating modes. The critical configurations $\tilde{\chi}=0$ and $\tilde{\chi}=\chi_{deg}$ coincide with the GR limit and the degenerate case, respectively. For $\tilde{\chi}=0$, indeed, we have $\chi_2=0$, so that no geometry-matter coupling is actually present at linear level. For $\tilde{\chi}=\chi_{deg}$, instead, both branches $\omega_{\pm}$ collapse to a unique dispersion relation where Landau damping can take place only for modes with $k^2>k_{0,deg}^2$, where $k_{0,deg}$ is now entirely determined by the thermodynamical parameters of the medium. From previous works on the theme \cite{Avelino:2012qe}, it is possible to put a constraint on the maximum value of $|\chi_2|$ from the Newtonian limit of the theory, by requesting that the modification induced on Poisson equation be negligible with respect to the experimental uncertainty. By considering this bound, together with assuming realistic values for the parameters characterizing the medium, wee see that $|\tilde{\chi}|\ll |\chi_{deg}|$ and the degenerate regime is actually not achievable. Moreover, the size of the upper bound $\chi_2^{max}$ determines that the interaction between dark matter and gravitational waves with frequencies around $100\, \text{Hz}$ is localized at very high redshifts, where the material medium is relativistic and the approximations leading to the analytic formulae for the dispersion relation and damping rate are not well grounded. Nonetheless, the possibility of damping for specific wavelengths cannot be ruled out in general. Indeed, by considering a much denser medium in non-relativistic conditions, the value of $\chi_{deg}$ can be made conveniently smaller, in order to gain access to other regimes, as the degenerate case $\tilde{\chi}=\chi_{deg}$. Another possibility is to investigate the dielectric function \eqref{eq: dielectric jutt} with fully numerical techniques, so that the assumption $x \ll 1$ can be omitted and the interaction with relativistic media, possibly leading to significant damping, can be properly described.       

\acknowledgments
The work of FB and GJO has been supported by grants PID2020-116567GB-C21, PID2023-149560NBC21 funded by MCIU/AEI/10.13039/501100011033 and FEDER, UE. 

\appendix
\section{Metric-affine formalism}\label{app: a}
\noindent In this appendix we review some basic notions about the metric-affine formalism we adopt throughout the paper. The Riemann tensor is defined in terms of the independent connection as:
\begin{equation}
    R\indices{^\rho_{\mu\sigma\nu}}=\partial_\sigma\Gamma\indices{^\rho_{\mu\nu}}-\partial_\nu\Gamma\indices{^\rho_{\mu\sigma}}+\Gamma\indices{^\rho_{\tau\sigma}}\Gamma\indices{^\tau_{\mu\nu}}-\Gamma\indices{^\rho_{\tau\nu}}\Gamma\indices{^\tau_{\mu\sigma}},
\end{equation}
and covariant derivatives act as
\begin{equation}
    \nabla_\mu T\indices{^\rho_\sigma}=\partial_\mu T\indices{^\rho_\sigma}+\Gamma\indices{^\rho_{\lambda\mu}}T\indices{^\lambda_\sigma}-\Gamma\indices{^\lambda_{\sigma\mu}}T\indices{^\rho_\lambda}.
\end{equation}
We are considering the affine connection as general as possible, so that it is possible to introduce the torsion and nonmetricity tensors, defined respectively by:
\begin{equation}
    \begin{split}        &t\indices{^\rho_{\mu\nu}}\equiv\Gamma\indices{^\rho_{\mu\nu}}-\Gamma\indices{^\rho_{\nu\mu}},\\
    &Q\indices{_{\rho\mu\nu}}\equiv-\nabla_\rho g_{\mu\nu}.
    \end{split}
\end{equation}
From the definition of non-metricity it is easy to see that the following holds $Q\indices{_{\rho}^{\mu\nu}}\equiv \nabla_\rho g^{\mu\nu}$.
In evaluating the equation of motion for the connection, it is useful to resort the generalized Palatini identity
\begin{equation}
    \delta R\indices{^\rho_{\mu\sigma\nu}}=\nabla_\sigma\delta\Gamma\indices{^\rho_{\mu\nu}}-\nabla_\nu\delta\Gamma\indices{^\rho_{\mu\sigma}}-t\indices{^\lambda_{\sigma\nu}}\delta\Gamma\indices{^\rho_{\mu\lambda}},
    \label{eq: pal identity}
\end{equation}
and the property for vector densities
\begin{equation}
    \int d^4 x\; \nabla_\mu\leri{\sqrt{-g} V^\mu}=\int d^4x\; \partial_\mu\leri{\sqrt{-g}V^\mu}+\int d^4x\; \sqrt{-g}\;t\indices{^\rho_{\mu\rho}} V^\mu.
    \label{eq: boundary term affine}
\end{equation}
Torsion and nonmetricity can be decomposed in their irreducible parts according the Lorentz representation, i.e.:
\begin{align}
    &t_{\mu\nu\rho} = \dfrac{1}{3}\left(t_{\nu}g_{\mu\rho}-t_{\rho}g_{\mu\nu}\right) +\dfrac{1}{6} \varepsilon_{\mu\nu\rho\sigma}S^{\sigma} + q_{\mu\nu\rho},\label{eq: torsion decomposition}\\
    &Q_{\rho\mu\nu}=\frac{5Q_\rho-2P_\rho}{18}g_{\mu\nu}-\frac{Q_{(\mu}g_{\nu)\rho}-4P_{(\mu}g_{\nu)\rho}}{9}+\Omega_{\rho\mu\nu}.
    \label{eq: non metricity decomposition}
\end{align}
For the torsion they are respectively the trace vector
\begin{equation}
t_{\mu} \equiv t \indices{^{\nu}_{\mu\nu}},
\end{equation}
the pseudotrace axial vector
\begin{equation}
S_{\mu} \equiv \varepsilon_{\mu\nu\rho\sigma}t^{\nu\rho\sigma},
\end{equation}
and the antisymmetric tensor $q_{\mu\nu\rho}=-q_{\mu\rho\nu}$ satisfying
\begin{equation}
\varepsilon^{\mu\nu\rho\sigma} q_{\nu\rho\sigma} = 0, \qquad q\indices{^{\mu}_{\nu\mu}} = 0.
\end{equation}
For what concerns the nonmetricity, we can identified instead the Weyl vector (first trace)
\begin{equation}
    Q_\rho=Q\indices{_\rho^\mu_\mu},
\end{equation}
the second trace
\begin{equation}
    P_\rho=Q\indices{^\mu_{\mu\rho}}=Q\indices{^\mu_{\rho\mu}},
\end{equation}
and the traceless part $\Omega_{\rho\mu\nu}$, obeying
\begin{equation}
    \Omega_{\rho\mu\nu}=\Omega_{\rho\nu\mu}.
\end{equation}
In terms of the Weyl vector it is possible to show that $\nabla_\rho\g=-\frac{\g Q_\rho}{2}$. \\ The affine connection can be further rewritten as
\begin{equation}
    \Gamma\indices{^\rho_{\mu\nu}}=L\indices{^\rho_{\mu\nu}}+N\indices{^\rho_{\mu\nu}}=L\indices{^\rho_{\mu\nu}}+K\indices{^\rho_{\mu\nu}}+D\indices{^\rho_{\mu\nu}},
    \label{eq: christoffel contorsion disformal}
\end{equation}
where $L\indices{^\rho_{\mu\nu}}$ denotes the Christoffel symbols and the contorsion and disformal tensors are given by, respectively
\begin{align}
    &K\indices{^\rho_{\mu\nu}}=\frac{1}{2}\leri{T\indices{^\rho_{\mu\nu}}-T\indices{_\mu^\rho_\nu}-T\indices{_\nu^\rho_\mu}}=-K\indices{_\mu^\rho_{\nu}},
    \label{eq: decomposition contorsion}\\
    &D\indices{^\rho_{\mu\nu}}=\frac{1}{2}\leri{Q\indices{_{\mu\nu}^\rho}+Q\indices{_{\nu\mu}^\rho}-Q\indices{^\rho_{\mu\nu}}}=D\indices{^\rho_{\nu\mu}}.
    \label{eq: decomposition disformal}
\end{align}
For a generic metric-affine structure the Riemann tensor is skew-symmetric only in its last two indices, so that we can in principle take the different traces
\begin{align}
    R_{\mu\nu}&\equiv R\indices{^\alpha_{\mu\alpha\nu}},\\
    \hat{R}_{\mu\nu}&\equiv R\indices{^\alpha_{\alpha\mu\nu}}=\partial_{[\mu}Q_{\nu]},\\
    \tilde{R}_{\mu\nu}&\equiv g_{\mu\tau}g^{\rho\sigma}R\indices{^\tau_{\rho\sigma\nu}},
\end{align}
which are called respectively Ricci, homothetic curvature and co-Ricci tensors. In terms of the distorsion tensor the Riemann curvature can be rewritten as
\begin{equation}
 R_{\mu\rho\nu\sigma}=\tensor[^{(L)}]{R}{_{\mu\rho\nu\sigma}}+\tensor[^{(L)}]{\nabla}{_{\nu}}N_{\mu\rho\sigma}-\tensor[^{(L)}]{\nabla}{_{\sigma}} N_{\mu\rho\nu}+N_{\mu\lambda\nu}N\indices{^\lambda_{\rho\sigma}}-N_{\mu\lambda\sigma}N\indices{^\lambda_{\rho\nu}},
    \label{eq: perturbative expansion riemann}
\end{equation}
where $\tensor[^{(L)}]{R}{^\mu_{\nu\rho\sigma}}$ and $\tensor[^{(L)}]{\nabla}{_{\mu}}$ are built from the Levi Civita connection. The affine Einstein tensor $G_{\mu\nu}$ can be therefore rewritten as
\begin{equation}
    G_{\mu\nu}=\tensor[^{(L)}]{G}{_{\mu\nu}}+A_{\mu\nu},
\end{equation}
where we introduced the tensor \begin{equation}
\begin{split}
    A_{\mu\nu}\equiv& \tensor[^{(L)}]{\nabla}{_{\rho}}N\indices{^\rho_{(\mu\nu)}}-\tensor[^{(L)}]{\nabla}{_{(\nu}}N\indices{^\rho_{\mu)\rho}}+N\indices{^\rho_{\lambda\rho}}N\indices{^\lambda_{(\mu\nu)}}-N\indices{^\rho_{\lambda(\nu}}N\indices{^\lambda_{\mu)\rho}}+\\
    &-\frac{1}{2}g_{\mu\nu}\leri{\tensor[^{(L)}]{\nabla}{_{\rho}}N\indices{^{\rho\sigma}_\sigma}-\tensor[^{(L)}]{\nabla}{_{\sigma}}N\indices{^{\rho\sigma}_{\rho}}+N\indices{^\rho_{\lambda\rho}}N\indices{^{\lambda\sigma}_\sigma}-N\indices{^\rho_{\lambda\sigma}}N\indices{^{\lambda\sigma}_\rho}}.
\end{split}
\label{eq: A tensor affine}
\end{equation}

\bibliographystyle{JHEP}
\bibliography{references}
\end{document}